\documentclass[aps,pre,twocolumn,superscriptaddress,longbibliography,nofootinbib,showkeys,showpacs]{revtex4-2}
\usepackage{graphicx}

\usepackage{xcolor}
\usepackage{mathtools,amssymb,amsmath}
\usepackage[colorlinks = true, citecolor=blue, linkcolor=black,urlcolor=purple]{hyperref}
\usepackage{cleveref}
\usepackage{braket}
\usepackage{enumitem}
\usepackage{xspace}

\crefname{equation}{Eq.}{Eqs.}
\crefname{figure}{Fig.}{Figs.}
\crefname{table}{Table}{Tables}
\crefname{section}{Sec.}{Secs.}
\crefname{appendix}{Appendix}{Appendices}

\Crefname{equation}{Equation}{Equations}
\Crefname{figure}{Figure}{Figures}
\Crefname{table}{Table}{Tables}
\Crefname{section}{Section}{Sections}
\Crefname{appendix}{Appendix}{Appendices}

\newcommand{\KP}{\hyperref[eq:KP]{\textsc{KP}}\xspace}
\newcommand{\QKP}{\hyperref[eq:QKP]{\textsc{Q-KP}}\xspace}
\newcommand{\EKP}[1][]{\hyperref[eq:KP_eq]{\textsc{EKP}\if\relax\detokenize{#1}\relax\else\ensuremath{(#1)}\fi}\xspace}
\newcommand{\QEKP}[1][]{\hyperref[eq:QKP_eq]{\textsc{Q-EKP}\if\relax\detokenize{#1}\relax\else\ensuremath{(#1)}\fi}\xspace}

\begin{document}

\makeatother
\title{Resource-efficient formulations of the Knapsack problem for quantum computing}

\author{Pim van den Heuvel}
\thanks{Corresponding author: Pim van den Heuvel}
\email{p.van.den.heuvel@fz-juelich.de}
\affiliation{J\"ulich Supercomputing Centre, Forschungszentrum J\"ulich, 52425 J\"ulich, Germany}
\affiliation{RWTH Aachen University, 52056 Aachen, Germany}
\author{Vrinda Mehta}
\affiliation{J\"ulich Supercomputing Centre, Forschungszentrum J\"ulich, 52425 J\"ulich, Germany}
\author{Dennis Willsch}
\affiliation{J\"ulich Supercomputing Centre, Forschungszentrum J\"ulich, 52425 J\"ulich, Germany}
\affiliation{FH Aachen University of Applied Sciences, 52066 Aachen, Germany}
\author{J. A. Monta\~nez-Barrera}
\affiliation{J\"ulich Supercomputing Centre, Forschungszentrum J\"ulich, 52425 J\"ulich, Germany}

\author{Kristel Michielsen}
\affiliation{J\"ulich Supercomputing Centre, Forschungszentrum J\"ulich, 52425 J\"ulich, Germany}
\affiliation{AIDAS, 52425 J\"ulich, Germany}
\affiliation{Faculty of Mathematics and Natural Sciences, University of Cologne, 50923 K\"oln, Germany}

\date{November 2026}

\begin{abstract}
Quantum computing has emerged as a new tool for solving combinatorial optimization problems. Quantum optimization approaches usually operate on the subclass of Quadratic Unconstrained Binary Optimization (QUBO) problems. Many optimization problems can be reformulated in this form, though there are often multiple ways to do so. The choice of formulation can considerably affect the performance of QUBO solvers in practice. In this work, we study QUBO formulations of the Knapsack problem, a canonical example of a problem with inequality constraints. In the standard formulation these constraints are converted into penalty terms, which requires additional slack variables. Here, we investigate three more resource-efficient approaches for formulating the Knapsack as a QUBO. The first two modify the standard formulation, either by deriving a smaller penalty constant or by reducing the number of slack variables. The third approach is a wholly different method, which avoids the use of slack variables entirely. All three approaches increase the probability of finding the optimal solution, both in simulations of the linear ramp Quantum Approximate Optimization Algorithm (LR-QAOA) and on a D-Wave quantum annealer. The slackless method performs best, increasing success probabilities by several orders of magnitude even for relatively small problem sizes.
\end{abstract}

\maketitle

\section{Introduction}\label{sec:Introduction}
Quantum computing provides a new avenue for solving combinatorial optimization problems. Various quantum optimization approaches have been developed in recent years, including algorithms for universal gate-based quantum computers, such as the Quantum Approximate Optimization Algorithm (QAOA) and its variants \cite{FarhQuantumApproximate2014,HadfQuantumApproximate2019,ZhouQuantumApproximate2020,MontlinearrampQAOA2025,BlekreviewQuantum2024}, as well as more specialized quantum devices, such as quantum annealers \cite{FarhQuantumAdiabatic2001,KadoQuantumannealing1998,JohnQuantumannealing2011,HaukPerspectivesquantum2020,YarkQuantumannealing2022}. The main problem class considered in quantum optimization is Quadratic Unconstrained Binary Optimization (QUBO). Many optimization problems can be (re)formulated as QUBO problems \cite{LucaIsingformulations2014}, and have been investigated using quantum optimization approaches, ranging from graph \cite{FarhQuantumApproximate2014,ZhouQuantumApproximate2020,HarrQuantumapproximate2021,HameExperimentalinvestigation2019,VodeAccuracyperformance2026} and constraint satisfaction problems \cite{BianSolvingSAT2020,PeloHighRoundQAOA2023,MehtQuantumannealing2022,MehtPerformancequantum2025} to scheduling and logistics \cite{VentQuantumAnnealing2016,CaruEvaluatingjob2022,Amarcasestudy2022,StolFlightGate2019,StolQuantumAnnealing2020} and financial portfolio optimization \cite{RoseSolvingOptimal2016,GranBenchmarkingQuantum2021,HodsPortfoliorebalancing2019,BranBenchmarkingperformance2022}.

The reformulation of a combinatorial optimization problem as a QUBO problem is generally not unique; multiple equally valid formulations may exist for the same problem.
These different QUBO formulations can differ in their structure and size (i.e., number of variables), and can therefore lead to differences in solver performance in practice \cite{RoseSolvingOptimal2016,FuchEfficientEncoding2021,TabiQuantumOptimization2020,HarwFormulatingSolving2021,MontUnbalancedpenalization2024,MontImprovingPerformance2023,Smittravellingsalesperson2025,DjidQuantumAnnealing2023,ChenPerformanceDomainWall2021}. This is particularly relevant for quantum optimization in the current noisy intermediate-scale quantum (NISQ) era \cite{PresQuantumComputing2018}, where quantum hardware remains limited in resources such as the number and quality of available qubits.
We remark that the choice of QUBO formulation can also affect the performance of classical QUBO solvers \cite{TamuPerformanceComparison2021,SekiBlackboxoptimization2022,CodoComparingQUBO2025}, although this is not the focus of our work. The effects of QUBO formulation on solver performance have also motivated research into standardizing and automating the process of converting optimization problems into QUBO form \cite{ZamaPyQUBOPython2022,DomiEncodingindependentoptimization2023,SawaEncodingtradeoffs2023,gawehn2026yael}.

Here, we study the QUBO formulation of the 0-1 Knapsack problem. A particular challenge in the QUBO formulation of the Knapsack problem is its inequality constraint. A standard approach to incorporate constraints into the unconstrained QUBO formulation is to include them in the problem's objective function as penalty terms \cite{LucaIsingformulations2014,Kochunconstrainedbinary2014,GlovQuantumBridge2019}. Penalty terms penalize variable assignments that violate the constraints; these terms are scaled by penalty constants. For inequality constraints in particular, additional slack variables are required to convert them into penalty terms. As each variable corresponds to one qubit in quantum optimization, this increases the number of required qubits.

Although the additional qubits make this standard slack-variable QUBO formulation for the Knapsack problem resource-intensive to solve, several studies have applied QAOA \cite{RochCrossEntropy2020} and quantum annealing \cite{PuseAdiabaticQuantum2020,RochCrossEntropy2021} to it. An alternative approach is to use an approximate penalty term that avoids the need for slack variables \cite{MontUnbalancedpenalization2024}. Other quantum algorithms have been applied to the Knapsack problem that handle the constraint without penalty terms, such as a variant of QAOA, dubbed copula-QAOA \cite{DamQuantumOptimization2021,MohsConstrainedQuantum2026}. Additional penalty-free algorithms include oracle-based methods \cite{GranKnapsackProblem2019,KeaLeveragingKnapsack2023,BuchPenaltyfreeapproach2025} and algorithms based on the quantum tree generator (QTG) \cite{Wilkquantumalgorithm2025,ChriQuantumTree2025}, which was designed specifically for the Knapsack problem.

In this paper, we investigate three approaches to improve the standard slack-variable QUBO formulation of the Knapsack problem:
\begin{enumerate}[label=(\roman*)]
    \item Using a smaller penalty constant
    \item Reducing the number of slack variables
    \item Eliminating slack variables altogether with a slackless method
\end{enumerate}
While the first two approaches modify the standard formulation, the third is a fundamentally different method. It solves a set of QUBO problems without slack variables, at least one of which yields the solution to the Knapsack problem. 

Smaller penalty constants have been reported to improve the performance of quantum approaches \cite{RochCrossEntropy2020,RochCrossEntropy2021,RochEffectPenalty2023,VillAnalyzingbehaviour2022,GuneQUBOFormulations2025,AlesAlleviatingquantum2025}, which motivates approach (i). To obtain a smaller penalty constant, we first introduce a general approach for deriving penalty constants, which is also applied in the slackless approach (iii) later. Similarly, approaches (ii) and (iii) are motivated by the aforementioned qubit overhead of the standard formulation: they reduce or eliminate the slack variables, which is also expected to improve solver performance.

We find that all three approaches increase the probability of finding the optimal solution for two example quantum optimizationsolvers, namely simulations of the QAOA variant LR-QAOA \cite{MontlinearrampQAOA2025}, and a D-Wave quantum annealer. The slackless approach (iii) achieves the largest improvement, with success probabilities that are several orders of magnitude higher than for the standard formulation, even for the relatively small problems that these solvers can currently handle.

Our paper is structured as follows. In \cref{sec:StandardForm}, we present the standard slack-variable QUBO formulation of the Knapsack problem and introduce the general approach for deriving sufficient penalty constants. In \cref{sec:ImprovedMult}, we derive a smaller penalty constant for this formulation (approach (i)). In \cref{sec:ReducedSlack}, we present approach (ii), which reduces the number of slack variables. In \cref{sec:SlacklessForm}, we discuss approach (iii), the slackless method. In \cref{sec:QuantumSolvers}, we describe the quantum solvers used in this work and specify our generation of Knapsack problem instances. We then compare the performance of the three approaches using these solvers in \cref{sec:Results}. Finally, we conclude in \cref{sec:Conclusion}.

\section{Knapsack problem formulation}\label{sec:StandardForm}

In this section, we first introduce the Knapsack problem in \cref{subsec:Knapsack}, before presenting its standard QUBO formulation in \cref{subsec:KnapsackQUBO}. Then, we present the general approach for deriving sufficient penalty constants in \cref{subsec:PenaltyConstant}.

\subsection{Knapsack problem}\label{subsec:Knapsack}

The Knapsack problem is a well-known NP-hard optimization problem, with applications in many fields \cite{KellKnapsackProblems2004}. It asks to maximize the total value of items carried inside a knapsack, while keeping the total weight of the knapsack below some threshold. Mathematically, we are given $n$ items with value $v_i > 0$ and weight $w_i > 0$ ($i \in \{1, 2, \ldots, n \}$) and a knapsack of capacity $W$. An instance \KP of the Knapsack problem is then formulated as
\phantomsection\label{eq:KP}
\begin{align}
    \text{Maximize} \quad & \sum_i^n v_i x_i \tag{\textsc{KP-obj}}, \label{eq:KP-obj} \\
    \text{subject to} \quad & \sum_i^n w_i x_i \leq W \tag{\textsc{KP-con}}, \label{eq:KP-con}
\end{align}
where the binary variables $x_i$ indicate whether an item is included in the knapsack ($x_i = 1$) or not ($x_i = 0$). Possible solutions are binary vectors $\mathbf{x} \in \{0, 1\}^n$ and an optimal solution to \KP is denoted $\mathbf{x}^{*}$. We restrict ourselves to instances with strictly positive integer weights, values, and capacities (any problem whose weights, values, and capacities are rational can, in any case, be converted into an equivalent integer problem by rescaling). We also assume non-trivial problems throughout the paper, meaning $w_i < W \; \forall \; i$ and $\sum_i w_i > W$. 

\subsection{Standard QUBO formulation}\label{subsec:KnapsackQUBO}

\begin{figure*}
    \centering
    \includegraphics[width=\textwidth]{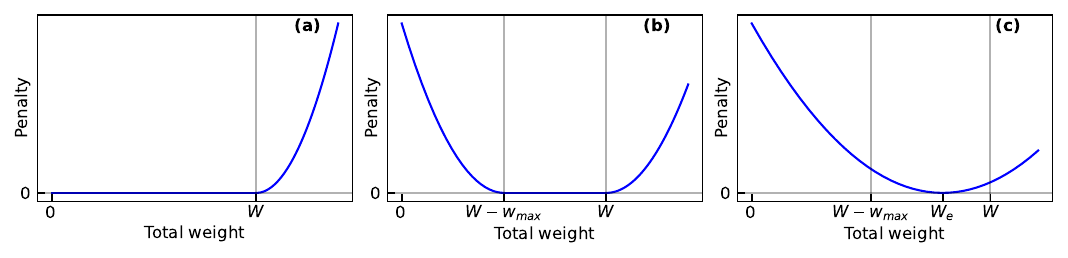}
    \caption{Schematic illustrations of the penalty terms in \QKP for (a) the full slack range $s \in \{0, 1, \dots, W \}$ (see \cref{subsec:KnapsackQUBO}) and (b) the reduced slack range $s \in \{0,1,\dots,w_{max}-1\}$ (see \cref{subsec:ReducedSlackRange}), and in \QEKP (c). The penalties are shown as functions of the total weight. In (a) and (b), the lowest achievable penalty over the slack variable range is shown for each total weight.
    }
    \label{fig:PenaltySketch}
\end{figure*}

To reformulate the Knapsack problem as a QUBO problem, the capacity constraint \eqref{eq:KP-con} is usually transformed to a penalty function through the introduction of the integer slack variable $s$:
\begin{equation}\label{eq:KP_penalty}
    P(\mathbf{x}, s) = \left(\sum_i^n w_i x_i - W + s\right)^2.
\end{equation}
By restricting the value of the slack variable to the knapsack capacity, i.e., $s \in \{0, 1, \dots, W \}$, any feasible solution that satisfies \eqref{eq:KP-con} can attain zero penalty by assigning $s = W - \sum_i w_i x_i$. In this case, $s$ intuitively represents the leftover, unused capacity of the knapsack. Infeasible solutions acquire a penalty that grows quadratically (for fixed $s$) with the amount of violation of the constraint, as shown in \cref{fig:PenaltySketch}(a). Using this penalty term, \KP can be reformulated as
\phantomsection\label{eq:QKP}
\begin{align*}
    \max_{\mathbf{x},\, s} f\left(\mathbf{x}, s\right) \tag{\textsc{Q-KP}}
\end{align*}
where
\begin{align}
    f\left(\mathbf{x}, s\right)
    &= \sum_i^n v_i x_i
    - \lambda_0 \left(\sum_i^n w_i x_i - W + s\right)^2.
    \tag{\textsc{Q-KP}}
\end{align}

This is the standard slack-variable QUBO formulation for the Knapsack problem, once the slack variable $s$ is expressed in terms of binary variables; we defer this to \cref{subsec:SlackEncoding} for ease of presentation. Moreover, the penalty constant $\lambda_0 > 0$ must be chosen such that infeasible solutions are sufficiently penalized and an optimal solution $\mathbf{x}^{*}$ to \KP is optimal to \QKP too (for some value of $s$), as illustrated in \cref{fig:PenaltyExplanation}.
Such a QUBO reformulation is referred to as \textit{valid} from here on. 
For each instance of \KP, there exists a minimum sufficient constant $\lambda_0^{\min}$ that still guarantees a \textit{valid} \QKP. However, since finding $\lambda_0^{\min}$ is generally an NP-hard problem itself \cite{AlesAlleviatingquantum2025}, $\lambda_0$ is typically chosen to be large enough to ensure a \textit{valid} reformulation across all instances.

A straightforward choice is $\lambda_0 > \sum_i v_i$. Since the penalty term takes integer values and is at least 1 for any infeasible solution, this ensures all infeasible solutions are penalized at least as much as the maximum total value they could obtain \cite{AyodPenaltyWeights2022, GarcExactsequential2022, HarwFormulatingSolving2021, AlesAlleviatingquantum2025}. 

A less conservative choice $\lambda_0 > v_{max}$, where $v_{max} = \max_{i} v_i$, is presented by \textcite{LucaIsingformulations2014}, though \textcite{QuinCharacterizingBenchmarking} show that this choice is only guaranteed to produce \textit{valid} reformulations when the slack variable $s$ is encoded into binary variables using the unary or binary representation (see \cref{subsec:SlackEncoding}). We reproduce this derivation below in a different form, without specifying any encoding for the slack variable yet, to highlight a general approach for deriving sufficient values for penalty constants. On the basis of this approach, we obtain an improved sufficient value for $\lambda_0$ in \cref{sec:ImprovedMult}, and we revisit this approach in the context of the slackless method in \cref{subsec:SlacklessConstants}.

\subsection{Penalty constant}\label{subsec:PenaltyConstant}

\begin{figure*}
    \centering
    \includegraphics[width=\textwidth]{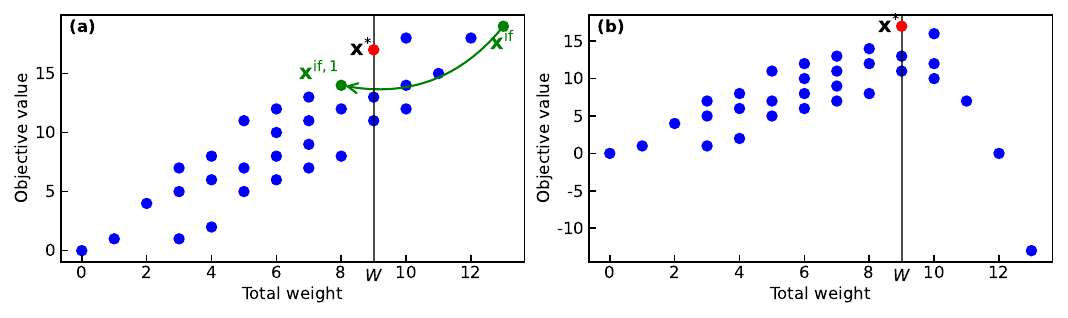}
    \caption{Objective values of all possible solutions to an example \KP instance with 5 items, plotted against their total weight, $\sum_i w_i x_i$. The optimal solution $\mathbf{x}^{*}$ to \KP is shown in red and the knapsack capacity $W$ is indicated by the vertical line. (a) Objective values obtained from the unpenalized objective function \eqref{eq:KP-obj}. The green arrow illustrates the construction of $\mathbf{x}^{\mathrm{if},1}$ from $\mathbf{x}^{\mathrm{if}}$ by removing a single item. (b) Objective values obtained from the penalized objective function \QKP. For each solution, only the maximum objective value over the slack variable range is shown. The penalty constant $\lambda_0$ is chosen sufficiently large here, such that the objective value of the optimal solution $\mathbf{x}^{*}$ to \KP is indeed larger than the objective value of any infeasible solution.}
    \label{fig:PenaltyExplanation}
\end{figure*}

A sufficient value of $\lambda_0$ is one for which the corresponding \QKP is a \textit{valid} QUBO reformulation of \KP, i.e., one for which $\mathbf{x}^{*}$ remains optimal to \QKP. For a sufficient $\lambda_0$, the following validity conditions must thus hold:
\begin{subequations}\label{eq:PenaltyConstantCorrectSlack}
\begin{align}
    \max_{s} f\left(\mathbf{x}^{*}, s \right) &\geq \max_{s} f\left(\mathbf{x}^{\mathrm{f}}, s\right) \quad \forall \; \mathbf{x}^{\mathrm{f}} \;,\label{eq:PenaltyConstantCorrectSlackFeas} \\
    \max_{s} f\left(\mathbf{x}^{*}, s \right) &\geq \max_{s} f\left(\mathbf{x}^{\mathrm{if}}, s\right) \quad \forall \; \mathbf{x}^{\mathrm{if}}\;.\label{eq:PenaltyConstantCorrectSlackInFeas}
\end{align}
\end{subequations}
Here, we partitioned the solutions into feasible solutions $\mathbf{x}^{\mathrm{f}} \in \{ \mathbf{x} \mid \sum_i w_i x_i \leq W \}$, which satisfy \eqref{eq:KP-con}, and infeasible solutions $\mathbf{x}^{\mathrm{if}} \in \{ \mathbf{x} \mid \sum_i w_i x_i > W \}$, which violate \eqref{eq:KP-con}. 

For all feasible solutions $\mathbf{x}^{\mathrm{f}}$ there exists a slack variable assignment for which the penalty term vanishes, namely when the slack is equal to the leftover weight, i.e., $s = W - \sum_i w_i x_i^{\mathrm{f}}$. So, $\max_{s} f\left(\mathbf{x}^{\mathrm{f}}, s\right) = \sum_i v_i x_i^{\mathrm{f}}$ and since $\mathbf{x}^{*}$ is feasible, and optimal, 
\begin{equation*}
    \max_{s} f\left(\mathbf{x}^{*}, s\right) = \sum_i v_i x_i^{*} \geq \sum_i v_i x_i^{\mathrm{f}} = \max_{s} f\left(\mathbf{x}^{\mathrm{f}}, s\right),
\end{equation*}
for all $\mathbf{x}^{\mathrm{f}}$. \Cref{eq:PenaltyConstantCorrectSlackFeas} is thus always satisfied, independent of the value of $\lambda_0$. 

Consequently, to derive a sufficient value for $\lambda_0$, we only need to consider \cref{eq:PenaltyConstantCorrectSlackInFeas}. 
Using the fact that infeasible solutions obtain minimum penalty when $s = 0$, we have $\max_{s} f\left(\mathbf{x}^{\mathrm{if}}, s\right) = \sum_i v_i x_i^{\mathrm{if}} - \lambda_0 P(\mathbf{x}^{\mathrm{if}}, 0)$, and \cref{eq:PenaltyConstantCorrectSlackInFeas} can be rewritten explicitly in terms of $\lambda_0$ as
\begin{equation}\label{eq:PenaltyConstantCorrectSlackInFeasExplicit}
    \sum_i v_i x_i^{*} \geq \sum_i v_i x_i^{\mathrm{if}} - \lambda_0 P(\mathbf{x}^{\mathrm{if}}, 0) \quad \forall \; \mathbf{x}^{\mathrm{if}}\;.
\end{equation}
Rearranging for $\lambda_0$ yields
\begin{equation}\label{eq:PenaltyConstantCorrectSlackInFeasExplicitRearranged}
    \lambda_0 \geq \frac{\sum_i v_i x_i^{\mathrm{if}} - \sum_i v_i x_i^{*}}{P(\mathbf{x}^{\mathrm{if}}, 0)} \quad \forall \; \mathbf{x}^{\mathrm{if}}\;,
\end{equation}
which is the condition for a sufficient value of $\lambda_0$, for which \QKP is a \textit{valid} reformulation of \KP. The minimum sufficient constant $\lambda_0^{\min}$ is the smallest value that satisfies \cref{eq:PenaltyConstantCorrectSlackInFeasExplicitRearranged} for all infeasible solutions $\mathbf{x}^{\mathrm{if}}$. It is given by the maximum of the right-hand side over all infeasible solutions:
\begin{equation}\label{eq:PenaltyConstantCorrectSlackInFeasExplicitRearrangedMax}
    \lambda_0^{\min} = \max_{\mathbf{x}^{\mathrm{if}}} \frac{\sum_i v_i x_i^{\mathrm{if}} - \sum_i v_i x_i^{*}}{P(\mathbf{x}^{\mathrm{if}}, 0)}.
\end{equation}

However, since the optimal solution $\mathbf{x}^{*}$ is not known a priori, \cref{eq:PenaltyConstantCorrectSlackInFeasExplicitRearrangedMax} cannot be evaluated directly (and this already suggests that finding $\lambda_0^{\min}$ is as difficult as finding the optimal solution itself). We thus seek a condition for a sufficient $\lambda_0$ that does not depend on $\mathbf{x}^{*}$. To this end, we introduce for each $\mathbf{x}^{\mathrm{if}}$ a (known) feasible counterpart $\mathbf{x}^{\mathrm{if},k}$, which is obtained by removing $k$ items from $\mathbf{x}^{\mathrm{if}}$, i.e., flipping $x_i$ for those items from $1$ to $0$ (see \cref{fig:PenaltyExplanation}(a)). Such a $\mathbf{x}^{\mathrm{if},k}$ always exists for some $k$, since it is always possible to remove all items from $\mathbf{x}^{\mathrm{if}}$, resulting in the empty solution, which is feasible. 

Since $\mathbf{x}^{\mathrm{if},k}$ is feasible, $\sum_i v_i x_i^{*} \geq \sum_i v_i x_i^{\mathrm{if},k}$, and the numerator of \cref{eq:PenaltyConstantCorrectSlackInFeasExplicitRearranged} can be upper bounded: $\sum_i v_i x_i^{\mathrm{if}} - \sum_i v_i x_i^{*} \leq \sum_i v_i x_i^{\mathrm{if}} - \sum_i v_i x_i^{\mathrm{if},k}$. Applying this upper bound to \cref{eq:PenaltyConstantCorrectSlackInFeasExplicitRearranged} yields the more conservative condition
\begin{equation}\label{eq:PenaltyConstantCorrectSlackInFeasExplicitRearrangedUpperBound}
    \lambda_0 \geq \frac{\sum_i v_i x_i^{\mathrm{if}} - \sum_i v_i x_i^{\mathrm{if},k}}{P(\mathbf{x}^{\mathrm{if}}, 0)} \quad \forall \; \mathbf{x}^{\mathrm{if}}\;.
\end{equation}
A sufficient value of $\lambda_0$ that does not depend on the unknown optimal solution $\mathbf{x}^{*}$ can then be obtained by taking the maximum of the right-hand side of \cref{eq:PenaltyConstantCorrectSlackInFeasExplicitRearrangedUpperBound}:
\begin{equation}\label{eq:Lambda0Hat}
    \hat{\lambda}_0^{(1)} = \max_{\mathbf{x}^{\mathrm{if}}} \frac{\sum_i v_i x_i^{\mathrm{if}} - \sum_i v_i x_i^{\mathrm{if},k}}{P(\mathbf{x}^{\mathrm{if}}, 0)}.
\end{equation}
The derivation of \cref{eq:Lambda0Hat} allows any choice of $\mathbf{x}^{\mathrm{if},k}$ for each $\mathbf{x}^{\mathrm{if}}$, so any such choice may also be used when computing $\hat{\lambda}_0^{(1)}$. Any value of $\hat{\lambda}_0^{(1)}$ obtained in this way is an upper bound on the minimal sufficient value: $\lambda_0^{\min} \leq \hat{\lambda}_0^{(1)}$. Alternatively, a potentially smaller value $\hat{\lambda}_0^{(2)}$ can be obtained by first minimizing over all $\mathbf{x}^{\mathrm{if},k}$ for each $\mathbf{x}^{\mathrm{if}}$:
\begin{equation}\label{eq:Lambda0Hat2}
    \hat{\lambda}_0^{(2)} = \max_{\mathbf{x}^{\mathrm{if}}} \frac{\min_{\mathbf{x}^{\mathrm{if},k}} \left(\sum_i v_i x_i^{\mathrm{if}} - \sum_i v_i x_i^{\mathrm{if},k}\right)}{P(\mathbf{x}^{\mathrm{if}}, 0)},
\end{equation}
which is also an upper bound on the minimal sufficient value: $\lambda_0^{\min} \leq \hat{\lambda}_0^{(2)} \leq \hat{\lambda}_0^{(1)}$.

Evaluating either \cref{eq:Lambda0Hat} or \cref{eq:Lambda0Hat2} is still computationally expensive, as it requires maximization over all $\mathbf{x}^{\mathrm{if}}$, which are generally not known without enumeration over all possible solutions. Instead, we can upper bound $\hat{\lambda}_0^{(1)}$ by bounding the numerator on its right-hand side from above and the denominator from below (the argument developed below also yields an upper bound on $\hat{\lambda}_0^{(2)}$).

In particular, we first partition the infeasible solutions $\mathbf{x}^{\mathrm{if}}$ into subsets $\mathcal{I}_k$, where each $\mathcal{I}_k$ contains those infeasible solutions for which the minimum number of items that must be removed from them to reach a feasible solution $\mathbf{x}^{\mathrm{if},k}$ is $k$. For each $\mathcal{I}_k$, we derive an upper bound on the ratio in \cref{eq:Lambda0Hat}; the largest of these bounds then serves as an overall upper bound on $\hat{\lambda}_0^{(1)}$.

We start with subset $\mathcal{I}_1$ to illustrate our argument. For each $\mathbf{x}^{\mathrm{if}} \in \mathcal{I}_1$, a feasible $\mathbf{x}^{\mathrm{if},1}$ can be constructed. The difference in total value between $\mathbf{x}^{\mathrm{if}}$ and this $\mathbf{x}^{\mathrm{if},1}$ is at most the maximum value of an item, $v_{max}$, which upper bounds the numerator of \cref{eq:Lambda0Hat}. The penalty in the denominator has a minimum value of $1$, as we consider integer-valued problems (see \cref{eq:KP_penalty}). Thus, the ratio in \cref{eq:Lambda0Hat} is at most $v_{max}$ for $\mathbf{x}^{\mathrm{if}} \in \mathcal{I}_1$. 

The argument for $\mathcal{I}_1$ extends to any arbitrary subset $\mathcal{I}_k$. For each $\mathbf{x}^{\mathrm{if}} \in \mathcal{I}_k$, there exists some $\mathbf{x}^{\mathrm{if},k}$, and the difference in total value between these solutions is at most $k v_{max}$, which again upper bounds the numerator of \cref{eq:Lambda0Hat}. The penalty in the denominator is the square of the excess weight $\sum_i w_i x_i^{\mathrm{if}} - W$ (see \cref{eq:KP_penalty}). For $\mathbf{x}^{\mathrm{if}} \in \mathcal{I}_k$, the excess weight is smallest when the items that are removed to obtain $\mathbf{x}^{\mathrm{if},k}$ have minimum weight $w_{min} = \min_i w_i$. Specifically, the smallest possible excess weight would occur when removing $k-1$ items with weight $w_{min}$ brought the total weight to the smallest infeasible value $W + 1$, before the final item is removed. The minimum excess weight for $\mathbf{x}^{\mathrm{if}} \in \mathcal{I}_k$ is thus $1 + (k-1) w_{min}$, and the penalty in the denominator of \cref{eq:Lambda0Hat} is at least $\left( 1 + (k-1) w_{min} \right)^2$. Since $w_{min} \geq 1$, $\left( 1 + (k-1) w_{min} \right)^2 \geq k^2$ (with equality only when $k=1$ or $w_{min}=1$), and the ratio in \cref{eq:Lambda0Hat} is thus at most $v_{max} / k$ for $\mathbf{x}^{\mathrm{if}} \in \mathcal{I}_k$.

The bound on the ratio in \cref{eq:Lambda0Hat} decreases with $k$ and the subset $\mathcal{I}_1$ thus yields the largest bound among all subsets $\mathcal{I}_k$. This is the overall upper bound on $\hat{\lambda}_0^{(1)}$, namely $\hat{\lambda}_0^{(1)} \leq v_{max}$. To guarantee that \QKP is a \textit{valid} reformulation of \KP,
\begin{equation}\label{eq:VmaxBound}
    \lambda_0 \geq v_{max},
\end{equation}
is then a sufficient value, confirming the result of \textcite{QuinCharacterizingBenchmarking}, without specifying any encoding for the slack variable $s$ yet. 

In practice, typically $\lambda_0 = (1 + \alpha) v_{max}$, for some small positive $\alpha$, is chosen, to avoid the possibility that the \QKP objective assigns the same value to some infeasible solution $\mathbf{x}^{\mathrm{if}}$ as to the optimal solution $\mathbf{x}^{*}$, as this could lead a (quantum) algorithm to return $\mathbf{x}^{\mathrm{if}}$ instead of $\mathbf{x}^{*}$.
That situation occurs if equality holds in \cref{eq:PenaltyConstantCorrectSlackInFeasExplicit} for some $\lambda_0$, $\mathbf{x}^{\mathrm{if}}$, and $\mathbf{x}^{*}$, and equivalently equality holds in \cref{eq:PenaltyConstantCorrectSlackInFeasExplicitRearranged}. 
Since we upper bounded the ratio in \cref{eq:Lambda0Hat}, and thereby the ratio in \cref{eq:PenaltyConstantCorrectSlackInFeasExplicitRearranged}, by $v_{max}$, equality in \cref{eq:PenaltyConstantCorrectSlackInFeasExplicitRearranged} requires $\lambda_0 \leq v_{max}$. Therefore, if $\lambda_0$ is chosen strictly larger than $v_{max}$, equality in \cref{eq:PenaltyConstantCorrectSlackInFeasExplicitRearranged} is not possible, and no $\mathbf{x}^{\mathrm{if}}$ can attain the same objective value as $\mathbf{x}^{*}$.

\section{Smaller penalty constant}\label{sec:ImprovedMult}
As mentioned before, there is evidence that smaller, but sufficient, penalty constants in QUBO formulations can improve the performance of quantum optimization approaches \cite{RochCrossEntropy2020,RochCrossEntropy2021,RochEffectPenalty2023,VillAnalyzingbehaviour2022,GuneQUBOFormulations2025,AlesAlleviatingquantum2025}. 
We show that, for \QKP, a sufficient value of $\lambda_0$ smaller than the usual choice in \cref{eq:VmaxBound}, i.e., $\lambda_0 \geq v_{\max}$, can be obtained with only polynomial overhead. Specifically, it is possible to derive a stricter upper bound on the numerator of \cref{eq:Lambda0Hat} using the solution found by a greedy algorithm. 

This value-first greedy algorithm sorts the items according to their value and iteratively inserts the highest-value item into the knapsack (when several items have equal value, it gives preference to items with smaller weight), terminating once adding the next item would exceed the knapsack capacity. Since the most computationally expensive step in this algorithm is sorting, it runs in log-linear time. We denote the set of items found by this value-first greedy algorithm as $\mathcal{G}_V$ and the value of the first item that would exceed the capacity (but is not inserted) as $v_s$. The items in the complementary set $\bar{\mathcal{G}}_V$\textemdash those items not included in the greedy solution\textemdash then have at most a value $v_s$, i.e., $v_i \leq v_s \; \forall \; i \in \bar{\mathcal{G}}_V$.

Every infeasible solution $\mathbf{x}^{\mathrm{if}}$ (with $\sum_i w_i x_i^{\mathrm{if}} > W$) must contain at least one item from $\bar{\mathcal{G}}_V$, since the total weight of items in $\mathcal{G}_V$ is not enough to exceed the knapsack capacity, i.e., $\sum_{i \in \mathcal{G}_V} w_i \leq W$. Moreover, if we remove from $\mathbf{x}^{\mathrm{if}}$ all items that are in $\bar{\mathcal{G}}_V$, the resulting solution contains only items from $\mathcal{G}_V$ (or no items) and is thus feasible. So it is always possible to construct $\mathbf{x}^{\mathrm{if},k}$ from $\mathbf{x}^{\mathrm{if}}$ by removing $k$ items from $\bar{\mathcal{G}}_V$ only. 

In the remainder of our reasoning to obtain the smaller penalty constant, we again upper bound the sufficient penalty constant $\hat{\lambda}_0^{(1)}$ (\cref{eq:Lambda0Hat}), following the same approach as in \cref{subsec:PenaltyConstant}. We partition the infeasible solutions $\mathbf{x}^{\mathrm{if}}$ into subsets $\mathcal{I}^V_k$, where each $\mathcal{I}^V_k$ again contains those infeasible solutions for which the minimum number of items that must be removed to reach $\mathbf{x}^{\mathrm{if},k}$ is $k$, now subject to the restriction that only items in $\bar{\mathcal{G}}_V$ may be removed. Considering first the subset $\mathcal{I}^V_1$, the numerator of \cref{eq:Lambda0Hat} can be bounded by the maximum value of items in $\bar{\mathcal{G}}_V$, $v_s$, instead of $v_{max}$. The ratio in \cref{eq:Lambda0Hat} is then at most $v_s$ for $\mathbf{x}^{\mathrm{if}} \in \mathcal{I}^V_1$. For an arbitrary subset $\mathcal{I}^V_k$, the same argument as presented in \cref{subsec:PenaltyConstant} can be followed to show that the ratio in \cref{eq:Lambda0Hat} is at most $v_s / k$ for $\mathbf{x}^{\mathrm{if}} \in \mathcal{I}^V_k$. As before, the largest bound among the subsets $\mathcal{I}^V_k$, the one for $k=1$, is the overall bound on $\hat{\lambda}_0^{(1)}$, namely $\hat{\lambda}_0^{(1)} \leq v_s$. The smaller sufficient value for the penalty constant is then finally: 
\begin{equation}\label{eq:VsBound}
    \lambda_0 \geq v_s,
\end{equation}
which guarantees that \QKP is a \textit{valid} reformulation of \KP. 

\section{Reduced slack range}\label{sec:ReducedSlack}

We first discuss how the slack variable $s$ in the \QKP formulation can be encoded into binary variables in \cref{subsec:SlackEncoding}, before explaining how the required number of binary slack variables can be reduced in \cref{subsec:ReduceSlack}.

\subsection{Encoding the slack variable}\label{subsec:SlackEncoding}
As mentioned in \cref{subsec:KnapsackQUBO}, \QKP is only a QUBO problem if the integer slack variable $s \in \{0, 1, \dots, W\}$ is converted into binary variables. There are several choices for this encoding (see \cite{DomiEncodingindependentoptimization2023,SawaEncodingtradeoffs2023} for an overview of them). 

Some encodings, such as the one-hot encoding, introduce additional constraints that must be enforced via additional penalty terms in the QUBO objective, each with its own penalty constant. Since the reasoning in \cref{subsec:PenaltyConstant} and \cref{sec:ImprovedMult} to derive sufficient values for the penalty constant is based on \QKP, which contains only a single penalty term, those values (\cref{eq:VmaxBound} and \cref{eq:VsBound}) may not be sufficient for any additional penalty constants. 
\textcite{QuinCharacterizingBenchmarking} showed that $\lambda_0 \geq v_{max}$ (\cref{eq:VmaxBound}) is indeed not a sufficient value for the additional penalty constant introduced by the one-hot encoding. 

Therefore, to use the sufficient values in \cref{eq:VmaxBound,eq:VsBound}, one should choose an encoding that does not introduce additional penalty terms, such as the unary or binary encoding. In this paper, we choose the binary encoding:
\begin{equation}\label{eq:slack_into_binary}
    s = \sum_{k=0}^{M - 1} 2^k s_k + \left(W - 2^{M} + 1\right)s_{M}\;,
\end{equation}
where $M = \lfloor \log(W) \rfloor$, so that a total of $M + 1$ binary variables $s_k$ are required. With this encoding, $s$ is restricted to the range $\{0, 1, \dots, W\}$, while different assignments of the $s_k$ may correspond to the same value of $s$. Since the slackless method (\cref{sec:SlacklessForm}) foregoes slack variables completely, the binary encoding was chosen for comparison, as it requires the fewest additional variables among unary and other common encodings without additional constraints \cite{DomiEncodingindependentoptimization2023}. 

\subsection{Reducing the number of slack variables}\label{subsec:ReduceSlack}
As was previously noted in the context of the Quadratic Knapsack problem \cite{PariAnalysisAcceleration2021,BontTranslatingConstraints2023,BorrNewQUBO2025}, a subtle property of the optimal solution to any \KP can be taken advantage of to lower the number of binary variables required to encode the slack variable $s$ in \QKP. Namely, the optimal solution's total weight $W^{*} = \sum_i w_i x_i^{*}$ must be larger than $W - w_{max}$, with $w_{max} = \max_i w_i$. This is because any solution with total weight at most $W - w_{max}$ can always be improved by inserting an additional item and hence cannot be optimal (since we chose $v_i > 0 \; \forall \; i$).

Accordingly, it is acceptable that the penalty term in \QKP (\cref{eq:KP_penalty}) assigns a penalty to solutions with total weight below $W - w_{max}$ (as well as to infeasible solutions), as those can never be optimal, as long as solutions with total weight in the range $\{W - w_{max} + 1, \dots, W\}$ receive no penalty (see \cref{fig:PenaltySketch}(b)). The slack variable in \QKP can be restricted to values up to $w_{max} - 1$ instead of $W$, i.e., $s \in \{0, 1, \dots, w_{max} - 1\}$, to obtain such a penalty function. The  required number of slack variables is then reduced; when using the binary encoding (\cref{eq:slack_into_binary}),  $\lfloor \log(w_{max} - 1) \rfloor + 1$ instead of $\lfloor \log(W) \rfloor + 1$ slack variables are required. 

In the same way, the number of required slack variables can be further reduced if a lower bound on the optimal solution's total weight tighter than $W^{*} > W - w_{max}$ is found. As we show in \cref{app:GreedyBoundProof}, the ratio-based greedy algorithm can be used to obtain such a bound. This greedy algorithm can be run in 
log-linear time, resulting in minimal overhead for the calculation of this bound. Even tighter bounds on the optimal solution's weight may exist, likely at the expense of an increased computational cost. We limit ourselves to the bound $W^{*} > W - w_{max}$ in the remainder of this paper for sake of simplicity.

\section{Slackless method}\label{sec:SlacklessForm}

We propose a method to reformulate the Knapsack problem as a set of QUBO problems without introducing additional slack variables. Each QUBO problem corresponds to an equality-constrained version of the Knapsack problem, where at least one of them is a \textit{valid} reformulation of the original problem. Similar ideas have been proposed in the context of the Quadratic Knapsack problem \cite{BorrNewQUBO2025} and the Bin Packing Problem \cite{DeAHybridquantumclassical2022,DeAComparativeBenchmark2022,XuDigitizedcounterdiabatic2025}.
In this section, we first introduce the equality-constrained version of \KP and its QUBO formulation in \cref{subsec:EqKnapsack}. Then we present our method, first for the equality-constrained problem in \cref{subsec:EKPForm}, and then, in \cref{subsec:QEKPForm}, for the QUBO reformulations of those problems. We refer to the latter as the slackless method. Sufficient penalty constants for the slackless method are derived in \cref{subsec:SlacklessConstants}. We detail the practical implementation of the slackless method in \cref{subsec:Implementation} and compare it against the equality-constrained method and the \QKP formulation in \cref{subsec:CompAndDisc}.

\subsection{Equality-constrained problem}\label{subsec:EqKnapsack}

The equality-constrained Knapsack problem \EKP[W_e] is a modification of the original \KP that replaces the inequality \eqref{eq:KP-con} with the equality \eqref{eq:EKP-con}:
\phantomsection\label{eq:KP_eq}
\begin{align}
    \text{Maximize} \quad & \sum_i^n v_i x_i 
    \tag{\textsc{EKP-obj}} \label{eq:EKP-obj} \\
    \text{subject to} \quad & \sum_i^n w_i x_i = W_e \;, \tag{\textsc{EKP-con}} \label{eq:EKP-con}
\end{align}
where $W_e \leq W$ is some (integer) feasible total weight. We denote the optimal solution to \EKP[W_e] by $\mathbf{x}^{*}_{W_e}$. The equality constraint \eqref{eq:EKP-con} can be converted into the quadratic penalty function
\begin{equation}\label{eq:KP_equalityPenalty}
    P_{\mathrm{eq}}(\mathbf{x}) = \left( \sum_i w_i x_i - W_e \right)^2 \;,
\end{equation}
which is shown in \cref{fig:PenaltySketch}(c). With this penalty, \EKP[W_e] can be reformulated as the QUBO problem \QEKP[W_e]
\phantomsection\label{eq:QKP_eq}
\begin{align*}
    \max_{\mathbf{x}} f_{\mathrm{eq}} \left(\mathbf{x}\right) \tag{\textsc{Q-EKP}}
\end{align*}
where
\begin{align}
    f_{\mathrm{eq}} \left(\mathbf{x}\right)
    &= \sum_i^n v_i x_i
    - \lambda_1 \left(\sum_i^n w_i x_i - W_e\right)^2.
    \tag{\textsc{Q-EKP}}
\end{align}
The penalty constant $\lambda_1 > 0$ must again be large enough for \QEKP[W_e] to be a \textit{valid} reformulation of \EKP[W_e]. As for the penalty constant of the QUBO reformulation \QKP of \KP (see \cref{subsec:KnapsackQUBO}), $\lambda_1 > \sum_i v_i$ would fulfill this requirement, since all infeasible solutions acquire a penalty at least as large as the maximum total value they can obtain. 

\subsection{Equality-constrained method}\label{subsec:EKPForm}

We now turn to our strategy for solving an instance \KP of the original problem.
Our method considers a set $\mathcal{E} = \{ \EKP[W_e] \mid W_e \in \mathcal{W} \}$ of equality-constrained versions of \KP. 
We choose $\mathcal{W} = \{ 1,2,\dots,W \}$, such that it covers all feasible (for \KP) total weights. If we consider an optimal solution $\mathbf{x}^{*}$ to \KP with total weight $W^{*} = w_i x_i^{*}$, this total weight then lies within this range, i.e., $W^{*} \in \mathcal{W}$.
The corresponding problem $\EKP[W^{*}] \in \mathcal{E}$ seeks to optimize the same objective function as \KP (cf. \eqref{eq:KP-obj} and \eqref{eq:EKP-obj}), albeit over the restricted set of feasible solutions $\{\mathbf{x} \mid \sum_i x_i w_i = W^{*} \}$, which includes the optimal solution $\mathbf{x}^{*}$ to \KP. Hence, an optimal solution $\mathbf{x}^{*}_{W^{*}}$ to \EKP[W^{*}] is equal to $\mathbf{x}^{*}$ and is thus optimal to \KP too (if \EKP[W^{*}] has multiple optimal solutions, all of them are optimal to \KP). 

Of course, $W^{*}$ is generally unknown before solving an instance \KP. 
Therefore, the entire range $\mathcal{W}$ must be considered; every $\EKP[W_e] \in \mathcal{E}$ must be solved to produce the corresponding set of solutions $\mathcal{X}^{*} = \{ \mathbf{x}^{*}_{W_e} \mid W_e \in \mathcal{W} \}$ (when no feasible solution exists for some \EKP[W_e], no solution is included in $\mathcal{X}^{*}$).
With our choice of $\mathcal{W}$, every $\mathbf{x}^{*}_{W_e} \in \mathcal{X}^{*}$ is feasible for \KP. The solution $\mathbf{x}^{*}_{W^{*}} \in \mathcal{X}^{*}$ is optimal to \KP, and can thus easily be identified from $\mathcal{X}^{*}$ as the solution yielding the maximum objective value. 

In \cref{subsec:ReduceSlack}, we exploited a property of the optimal solution to any $\KP$ to derive a lower bound on the total weight of the optimal solution: $W^{*} > W - w_{max}$. In addition to enabling a reduction in the number of slack variables, this bound allows us to restrict the range of possible total weights considered to $\mathcal{W} = \{ W - w_{max} + 1, W - w_{max} + 2, \dots, W \}$, while still ensuring that $W^{*} \in \mathcal{W}$. As a result, there are only $w_{max}$ (instead of $W$) problems in $\mathcal{E}$ that need to be solved.

\subsection{Outline of the slackless method}\label{subsec:QEKPForm}

For the slackless method, the \EKP problems in the method developed in \cref{subsec:EKPForm} are replaced by \QEKP problems. Specifically, we replace the set $\mathcal{E}$ by the corresponding set of QUBO problems $\mathcal{Q} = \{ \QEKP[W_e] \mid W_e \in \mathcal{W} \}$, which includes the problem \QEKP[W^{*}]. As long as this problem is a \textit{valid} reformulation of \EKP[W^{*}], an optimal solution $\mathbf{x}^{*}_{W^{*}}$ to \EKP[W^{*}], which is optimal to \KP, is optimal to \QEKP[W^{*}] too. Solving \QEKP[W^{*}] thus solves the original problem \KP. 

Since $W^{*}$ is not known a priori, every problem $\QEKP[W_e] \in \mathcal{Q}$ must still be solved. Additionally, a sufficient value for the penalty constant $\lambda_1$ is required, which ensures that $\QEKP[W^{*}]$ is indeed a \textit{valid} reformulation of \EKP[W^{*}]. Two expressions for such a sufficient value of $\lambda_1$ are derived in \cref{subsec:SlacklessConstants}. 
These expressions are then used to determine the penalty constants for all problems in $\mathcal{Q}$, although they are guaranteed to yield sufficient values only for the problem(s) with $W_e = W^{*}$.
Further details on the implementation of the slackless method are given in \cref{subsec:Implementation}. 

\subsection{Penalty constants}\label{subsec:SlacklessConstants}

We present two ways to derive a sufficient value for the penalty constant $\lambda_1$ for \QEKP[W^{*}], initially proceeding analogously to the derivation of a sufficient penalty constant for \QKP (see \cref{subsec:PenaltyConstant}).
Recall that a sufficient value of $\lambda_1$ is one for which \QEKP[W^{*}] is a \textit{valid} reformulation of \EKP[W^{*}], i.e., for which $\mathbf{x}^{*}_{W^{*}}$ is optimal to \QEKP[W^{*}]. 
Since $\mathbf{x}^{*}_{W^{*}}$ is optimal to \KP too, we denote it simply by $\mathbf{x}^{*}$ in the remainder of this derivation, where $\mathbf{x}^{*}$ is understood to have total weight exactly $W^{*}$.
A sufficient value of $\lambda_1$ is then one that ensures the following conditions:
\begin{subequations}\label{eq:PenaltyConstantCorrectEqual}
\begin{align}
    f_{\mathrm{eq}} \left( \mathbf{x}^{*} \right) &\geq f_{\mathrm{eq}} \left( \mathbf{x}^{\mathrm{f}} \right) \quad \forall \; \mathbf{x}^{\mathrm{f}}\;, \label{eq:PenaltyConstantCorrectEqualFeas} \\
    f_{\mathrm{eq}} \left( \mathbf{x}^{*} \right) &\geq f_{\mathrm{eq}} \left( \mathbf{x}^{\mathrm{if}} \right) \quad \forall \; \mathbf{x}^{\mathrm{if}}\;,\label{eq:PenaltyConstantCorrectEqualInFeas}        
\end{align}
\end{subequations}
where we partitioned the solutions into feasible solutions $\mathbf{x}^{\mathrm{f}}$ and infeasible solutions $\mathbf{x}^{\mathrm{if}}$. Here, a feasible solution $\mathbf{x}^{\mathrm{f}}$ is defined with respect to the constraint \eqref{eq:KP-con} of the original problem \KP, that is, $\mathbf{x}^{\mathrm{f}} \in \{ \mathbf{x} \mid \sum_i w_i x_i \leq W \}$ and $\mathbf{x}^{\mathrm{if}} \in \{ \mathbf{x} \mid \sum_i w_i x_i > W \}$.

The minimum value of the penalty term (\cref{eq:KP_equalityPenalty}) for \QEKP[W^{*}] is zero, so for any solution $\mathbf{x}$ it holds that $f_{\mathrm{eq}} \left( \mathbf{x} \right) \leq \sum_i v_i x_i$. In particular, for the optimal solution $\mathbf{x}^{*}$ with total weight $W^{*}$ the penalty term is zero: $f_{\mathrm{eq}} \left( \mathbf{x}^{*} \right) = \sum_i v_i x_i^{*}$. 
Therefore:
\begin{equation}\label{eq:PenaltyConstantCorrectEqualFeasProof}
    f_{\mathrm{eq}} \left( \mathbf{x}^{*} \right) = \sum_i v_i x_i^{*} \geq \sum_i v_i x^{\mathrm{f}}_i \geq f_{\mathrm{eq}} \left( \mathbf{x}^{\mathrm{f}} \right) \quad \forall \; \mathbf{x}^{\mathrm{f}}\;,
\end{equation}
where the first inequality follows from the optimality of $\mathbf{x}^{*}$ to \KP. \Cref{eq:PenaltyConstantCorrectEqualFeas} thus holds independently of the value of $\lambda_1$.

Hence, only \cref{eq:PenaltyConstantCorrectEqualInFeas} remains to derive a sufficient value for $\lambda_1$. By explicitly expressing $f_{\mathrm{eq}} \left( \mathbf{x}^{\mathrm{if}} \right) = \sum_i v_i x_i^{\mathrm{if}} - \lambda_1 P_{\mathrm{eq}} \left( \mathbf{x}^{\mathrm{if}} \right)$, it becomes:
\begin{equation}\label{eq:ExplicitPenaltyConstantEqual}
    \sum_i v_i x_i^{*} \geq  \sum_i v_i x_i^{\mathrm{if}} - \lambda_1 P_{\mathrm{eq}} \left( \mathbf{x}^{\mathrm{if}} \right) \quad \forall \; \mathbf{x}^{\mathrm{if}}\;,
\end{equation}
which is rearranged to yield the condition for a sufficient $\lambda_1$:
\begin{equation}\label{eq:Lambda1BoundOptimal}
    \lambda_1 \geq \frac{\sum_i v_i x_i^{\mathrm{if}} - \sum_i v_i x_i^{*}}{P_{\mathrm{eq}} \left( \mathbf{x}^{\mathrm{if}} \right)} \quad \forall \; \mathbf{x}^{\mathrm{if}}\;.
\end{equation}
Taking the maximum of the right-hand side of \cref{eq:Lambda1BoundOptimal} over all $\mathbf{x}^{\mathrm{if}}$ would give the smallest sufficient constant $\lambda_1^{\min}$ (as was the case for \cref{eq:PenaltyConstantCorrectSlackInFeasExplicitRearranged}). 

To obtain a practically computable sufficient condition for $\lambda_1$, we reuse the notation $\mathbf{x}^{\mathrm{if},k}$ to denote a feasible solution (with respect to \KP) obtained from an infeasible solution $\mathbf{x}^{\mathrm{if}}$ by removing $k$ items (see \cref{subsec:PenaltyConstant}). Since $\sum_i v_i x_i^{*} \geq \sum_i v_i x_i^{\mathrm{if},k}$ for any $\mathbf{x}^{\mathrm{if},k}$ (\cref{eq:PenaltyConstantCorrectEqualFeasProof}), the numerator of \cref{eq:Lambda1BoundOptimal} can be upper bounded: $\sum_i v_i x_i^{\mathrm{if}} - \sum_i v_i x_i^{*} \leq \sum_i v_i x_i^{\mathrm{if}} - \sum_i v_i x_i^{\mathrm{if},k}$. Applying this upper bound gives a more conservative condition on the sufficient value:
\begin{equation}\label{eq:Lambda1BoundK}
    \lambda_1 \geq  \frac{\sum_i v_i x_i^{\mathrm{if}} - \sum_i v_i x_i^{\mathrm{if},k}}{P_{\mathrm{eq}} \left( \mathbf{x}^{\mathrm{if}} \right)} \quad \forall \; \mathbf{x}^{\mathrm{if}}.
\end{equation}

Continuing analogously to \cref{subsec:PenaltyConstant}, a sufficient value can be obtained from \cref{eq:Lambda1BoundK} by taking the maximum of its right-hand side over all $\mathbf{x}^{\mathrm{if}}$: 
\begin{equation}\label{eq:Lambda1Hat}
    \hat{\lambda}_1^{(1)} = \max_{\mathbf{x}^{\mathrm{if}}} \frac{\sum_i v_i x_i^{\mathrm{if}} - \sum_i v_i x_i^{\mathrm{if},k}}{P_{\mathrm{eq}} \left( \mathbf{x}^{\mathrm{if}} \right)},
\end{equation}
where any choice of $\mathbf{x}^{\mathrm{if},k}$ can be used for each $\mathbf{x}^{\mathrm{if}}$ to compute $\hat{\lambda}_1^{(1)}$ and for any value so obtained it holds that $\lambda_1^{\min} \leq \hat{\lambda}_1^{(1)}$. We now proceed to find two upper bounds on $\hat{\lambda}_1^{(1)}$. 

For the first bound, we consider again the value-first greedy algorithm introduced in \cref{sec:ImprovedMult}, which iteratively selects the highest-value items that fit within the knapsack. It outputs the set of selected items $\mathcal{G}_V$ and the value $v_s$ of the first item that did not fit. The value of items in the complementary set $\bar{\mathcal{G}}_V$ is maximally $v_s$: $v_i \leq v_s \; \forall \; i \in \bar{\mathcal{G}}_V$. 
In \cref{sec:ImprovedMult} it was shown that it is always possible to construct $\mathbf{x}^{\mathrm{if},k}$ from $\mathbf{x}^{\mathrm{if}}$ by removing items that are in $\bar{\mathcal{G}}_V$ only. 

We again partition the infeasible solutions $\mathbf{x}^{\mathrm{if}}$ into subsets $\mathcal{I}^V_k$, as explained in \cref{sec:ImprovedMult}. For each $\mathcal{I}^V_k$, we derive an upper bound on the ratio in \cref{eq:Lambda1Hat}. The largest of these bounds is then the overall upper bound on $\hat{\lambda}_1^{(1)}$.

For $\mathbf{x}^{\mathrm{if}} \in \mathcal{I}^V_k$, the difference in total value in the numerator of \cref{eq:Lambda1Hat} is at most $k v_s$.
The penalty in the denominator of \cref{eq:Lambda1Hat} is $\left( \sum_i w_i x_i - W^{*} \right)^2 $ (see \cref{eq:KP_equalityPenalty}) for \QEKP[W^{*}]. Following the same reasoning as in \cref{subsec:PenaltyConstant}, the smallest possible excess weight $\sum_i w_i x_i^{\mathrm{if}} - W$ for $\mathbf{x}^{\mathrm{if}} \in \mathcal{I}^V_k$ would occur when removing $k - 1$ items with weight $w_{min} \left( \bar{\mathcal{G}}_V \right) = \min_{i \in \bar{\mathcal{G}}_V} w_i$ brought the total weight to the smallest infeasible total weight $W + 1$.
This leads to a minimum penalty for $\mathbf{x}^{\mathrm{if}} \in \mathcal{I}^V_k$ of $\left( U + 1 + (k-1) w_{min} \left( \bar{\mathcal{G}}_V \right) \right)^2$, where we define the unused capacity as $U = W - W^{*}$. 

Together, these results yield a collection of upper bounds on the ratio in \cref{eq:Lambda1Hat}, one for each subset $\mathcal{I}^V_k$. The overall upper bound on $\hat{\lambda}_1^{(1)}$ is given by the maximum over these bounds:
\begin{equation}\label{eq:Lambda1BoundValues}
    \hat{\lambda}_1^{(1)} \leq \max_k \frac{k v_s}{\left( U + 1 + (k-1) w_{min} \left( \bar{\mathcal{G}}_V \right) \right)^2}.
\end{equation}
If $U > 2 w_{min} \left( \bar{\mathcal{G}}_V \right) - 1$, the maximum of the ratio in \cref{eq:Lambda1BoundValues} is attained for either $k = \left\lfloor \left( U + 1 - w_{min} \left( \bar{\mathcal{G}}_V \right) \right) / w_{min} \left( \bar{\mathcal{G}}_V \right) \right\rfloor$ or $k = \left\lceil \left( U + 1 - w_{min} \left( \bar{\mathcal{G}}_V \right) \right) / w_{min} \left( \bar{\mathcal{G}}_V \right) \right\rceil$. If $U \leq 2 w_{min} \left( \bar{\mathcal{G}}_V \right) - 1$, the maximum is instead attained at $k = 1$. 
Choosing $\lambda_1$ larger than the bound on $\hat{\lambda}_1^{(1)}$ in \cref{eq:Lambda1BoundValues} ensures that it is sufficient, and guarantees that \QEKP[W^{*}] is a \textit{valid} reformulation of \EKP[W^{*}].

The second bound on $\hat{\lambda}_1^{(1)}$ is derived analogously to the first, but is based on the weight-first greedy algorithm rather than the value-first. This algorithm iteratively inserts the lowest-weight items into the knapsack (in case of equal-weight items, it prioritizes items with higher value), and stops when adding the next item would exceed the knapsack capacity. It returns the set of selected items $\mathcal{G}_W$ and the weight $w_s$ of the first item that did not fit. The complementary set $\bar{\mathcal{G}}_W$ then contains items with weight at least $w_s$: $w_i \geq w_s \; \forall \; i \in \bar{\mathcal{G}}_W$. Following the same argument as in \cref{sec:ImprovedMult}, $\mathbf{x}^{\mathrm{if},k}$ can always be constructed from $\mathbf{x}^{\mathrm{if}}$ by removing items from $\bar{\mathcal{G}}_W$ only. 

We partition the $\mathbf{x}^{\mathrm{if}}$ into subsets $\mathcal{I}^W_k$ as in \cref{sec:ImprovedMult}, now restricting the items that are removed to $\bar{\mathcal{G}}_W$. We derive an upper bound on the ratio in \cref{eq:Lambda1Hat} for each $\mathcal{I}^W_k$, and the maximum over these is the second upper bound on $\hat{\lambda}_1^{(1)}$.

The numerator of \cref{eq:Lambda1Hat} can be bounded by $k v_{max} \left( \bar{\mathcal{G}}_W \right)$, where $v_{max} \left( \bar{\mathcal{G}}_W \right) = \max_{i \in \bar{\mathcal{G}}_W} v_i$, for $\mathbf{x}^{\mathrm{if}} \in \mathcal{I}^W_k$. Using the same reasoning as for the first bound, the denominator is at least $\left( U + 1 + (k-1) w_s \right)^2$. The second upper bound on $\hat{\lambda}_1^{(1)}$ is then the maximum over the bounds for each $\mathcal{I}^W_k$:
\begin{equation}\label{eq:Lambda1BoundWeights}
    \hat{\lambda}_1^{(1)} \leq \max_k  \frac{k v_{max} \left( \bar{\mathcal{G}}_W \right)}{\left( U + 1 + (k-1) w_s \right)^2}\;.
\end{equation}

As for the first bound, the maximum in \cref{eq:Lambda1BoundWeights} is reached at $k=1$ if $U \leq 2w_s - 1$. Combining properties of the optimal and the weight-first greedy solution, it can be shown that this condition on $U$ always holds.
Since the weight-first greedy algorithm includes items in increasing order of weight, $w_s$ is the \textit{maximum} weight for which the set of items $\mathcal{T}(w_s - 1) = \{ i \mid w_i \leq w_s - 1 \}$ fits within the knapsack, i.e., for which $\sum_{i \in \mathcal{T}(w_s - 1)} w_i \leq W$. 
Any $w$ for which $\mathcal{T}(w)$ fits within the knapsack is thus at most $w_s - 1$. We show next that this is the case for the unused capacity $U$. 

The set of items $\mathcal{T}(U) = \{ i \mid w_i \leq U \}$ must be included in an optimal solution $\mathbf{x}^{*}$ with total weight $W^{*} = W - U$. Indeed, if any item $i \in \mathcal{T}(U)$ were not included in the optimal solution, the total value could be increased by adding this item, and the solution would not be optimal. Consequently, $\mathcal{T}(U)$ fits within the knapsack, as $\sum_{i \in \mathcal{T}(U)} w_i \leq W^{*} \leq W$. Thus, $U \leq w_s - 1$, and the condition $U \leq 2w_s - 1$ is always satisfied. The maximum in \cref{eq:Lambda1BoundWeights} is therefore always attained at $k=1$, so \cref{eq:Lambda1BoundWeights} becomes
\begin{equation}\label{eq:Lambda1BoundWeightFinal}
    \hat{\lambda}_1^{(1)} \leq \frac{v_{max} \left( \bar{\mathcal{G}}_W \right)}{\left( U + 1 \right)^2},
\end{equation}
and setting $\lambda_1$ to at least this bound is sufficient to ensure that \QEKP[W^{*}] is a \textit{valid} reformulation of \EKP[W^{*}].

Finally, we note that it is also possible to bound \cref{eq:Lambda1Hat} without the use of greedy algorithms, as in \cref{subsec:PenaltyConstant}. Specifically, if the $\mathbf{x}^{\mathrm{if}}$ are partitioned into subsets $\mathcal{I}_k$ (see \cref{subsec:PenaltyConstant}), then, for $\mathbf{x}^{\mathrm{if}} \in \mathcal{I}_k$, the numerator in \cref{eq:Lambda1Hat} can be upper bounded by $k v_{max}$, while the denominator can be lower bounded by $\left( U + 1 + (k-1) w_{min} \right)^2$. Combining these expressions yields an analogous bound to \cref{eq:Lambda1BoundValues} and \cref{eq:Lambda1BoundWeights}; however, the latter two bounds are tighter. 

\subsection{Implementation of the slackless method}\label{subsec:Implementation}

We use the restricted range of total weights $\mathcal{W} = \{ W - w_{max} + 1, W - w_{max} + 2, \dots, W \}$ (that includes $W^{*}$ still) to construct the set of problems $\mathcal{Q} = \{ \QEKP[W_e] \mid W_e \in \mathcal{W} \}$ for the slackless method. The bounds in \cref{eq:Lambda1BoundValues,eq:Lambda1BoundWeightFinal} on a sufficient value of $\lambda_1$ are derived specifically for the problem \QEKP[W^{*}]. However, it is unknown which problem $\QEKP[W_e] \in \mathcal{Q}$ has $W_e = W^{*}$. We therefore compute these bounds for each problem $\QEKP[W_e] \in \mathcal{Q}$, using the corresponding total weight $W_e$ (that is, we take $W_e$ as the value of $W^{*}$ in \cref{eq:Lambda1BoundValues,eq:Lambda1BoundWeightFinal}), and select the smaller of the two bounds as the penalty constant $\lambda_1$ for that problem. For those \QEKP[W_e] with $W_e = W^{*}$, the penalty constant is then sufficient, and their optimal solutions will be optimal to the original problem \KP too.

For the other $\QEKP[W_e] \in \mathcal{Q}$ with $W_e \neq W^{*}$, the penalty constant may not be sufficient; those problems may not be \textit{valid} reformulations of their respective \EKP[W_e]. Nevertheless, all problems $\QEKP[W_e] \in \mathcal{Q}$ must still be solved, as $W^{*}$ is unknown a priori. This yields a set of solutions, from which the solution $\mathbf{x}^{*}_{W^{*}}$ produced by \QEKP[W^{*}], which is optimal to \KP, can be identified as the feasible solution with maximum objective value. If, in practice, a solver does not return the optimal solution for some of the $\QEKP[W_e] \in \mathcal{Q}$, the best solution can still be selected in the same way among all returned solutions.

\subsection{Comparison and discussion}\label{subsec:CompAndDisc}

As opposed to the problem $\EKP[W_e] \in \mathcal{E}$ (see \cref{subsec:EKPForm}), whose solution space is explicitly limited to the feasible set $\{\mathbf{x} \mid \sum_i x_i w_i = W_e \}$, the problem $\QEKP[W_e] \in \mathcal{Q}$ allows any solution $\mathbf{x} \in \{ 0, 1 \} ^n$ (but penalizes infeasible ones). This also means that, whenever no solution with total weight $W_e$ exists, \EKP[W_e] has an empty feasible set, while \QEKP[W_e] still has an optimal solution, albeit one with total weight different from $W_e$.

Furthermore, since \QEKP[W_e] may not be a \textit{valid} reformulation of \EKP[W_e] when $W_e \neq W^{*}$, the optimal solution to \QEKP[W_e] need not have total weight equal to $W_e$ in any case. It may even be a solution that is optimal to \KP with total weight $W^{*}\neq W_e$. An example of such a case is shown in \cref{fig:SlacklessMethodsQAOA}(b) below. More generally, it is also possible that \QEKP[W_e] assigns a near-optimal objective value to an optimal solution $\mathbf{x}^{*}$ to \KP. In such cases, solvers which are expected to sample near-optimal and optimal solutions, such as the quantum solvers used in this work (see \cref{sec:QuantumSolvers}), may still return this optimal solution $\mathbf{x}^{*}$ when solving \QEKP[W_e]. 

Unlike the conventional QUBO formulation \QKP, the QUBO problems $\QEKP \in \mathcal{Q}$ do not require the use of slack variables. Instead, the slackless method requires solving all $w_{max}$ problems in $\mathcal{Q}$. The slackless method thus trades the approximately $\log(w_{max})$ additional slack qubits required by the conventional \QKP formulation (see \cref{subsec:ReduceSlack}) for $w_{max}$ executions of the QUBO solver of choice. 

Suppose a solver has exponential runtime $O(2^{N})$ in the number of variables $N$. Then, the runtime for the \QKP formulation with roughly $N = n + \log(w_{max})$ variables is $O(2^{n + \log(w_{max})}) = O(w_{max} 2^n)$. Solving all $w_{max}$ slackless problems in $\mathcal{Q}$, each with $N = n$ variables, likewise yields an asymptotic total runtime of $O(w_{max} 2^n)$, when using the same solver. A simple example of a solver with runtime $O(2^{N})$ is exhaustive search. On the other hand, the quantum solvers used in this work (see \cref{sec:QuantumSolvers}) are heuristic solvers that do not guarantee finding an optimal solution, but have the potential of running in less time. Nevertheless, in practice, their runtime has been shown to be exponential in the number of variables for a range of problems \cite{Manddeceptivestep2018,ShayEvidencescaling2024,WillGPUacceleratedsimulations2022,statefactoringquantumcomputers,MehtQuantumannealing2022,Mehthardnessquadratic2022,RobeSimonsPeriod2025}.

The computational cost for calculating the bounds in \cref{eq:Lambda1BoundValues,eq:Lambda1BoundWeightFinal} does not change the exponential runtime. For a single problem $\QEKP[W_e] \in \mathcal{Q}$, the runtime for this procedure is polynomial in the size $n$ of the original problem $\KP$, since the sorting steps in both greedy algorithms are the most expensive operations with respect to $n$. Since there are $|\mathcal{Q}|=w_{max}$ problems, the runtime for determining all penalty constants scales linearly with $w_{max}$. The greedy algorithms need to be executed only once, however; their results can be reused for all evaluations of \cref{eq:Lambda1BoundValues,eq:Lambda1BoundWeightFinal}.

\section{Quantum solvers}\label{sec:QuantumSolvers}
We employ two different quantum solvers to solve the proposed QUBO formulations of the Knapsack problem. Specifically, we consider simulations of the linear ramp Quantum Approximate Optimization Algorithm (LR-QAOA) and the D-Wave Advantage2 System JUPSI located at Forschungszentrum J\"{u}lich.

Both solvers represent the QUBO problem by an Ising Hamiltonian that is a combination of Pauli-$z$ operators $\sigma_i^z$. To transform the binary descriptions of the QUBO problems to this Ising form, the binary variables $x_i \in \{0,1\}$ are first mapped to spin variables according to the transformation $x_i \to (1-\sigma_i)/2$, where $\sigma_i \in \{-1,1\}$ are the spin variables. The spin variables are then represented by Pauli-$z$ operators $\sigma_i^z$ in the problem Hamiltonian. 

\subsection{Linear ramp Quantum Approximate Optimization Algorithm}\label{subsec:LR-QAOA}
The Quantum Approximate Optimization Algorithm (QAOA) is a well-known gate-based algorithm for solving QUBO problems \cite{FarhQuantumApproximate2014}. Its circuit consists of alternating unitaries generated by the problem Hamiltonian $H_P$ and a driver Hamiltonian $H_D$, which is typically chosen to be the transverse-field Hamiltonian $H_D = - \sum_{i=1}^{n}\sigma_i^x$. The problem Hamiltonian represents the QUBO problem in its Ising form: 
\begin{equation}\label{eq:ProblemHam}
H_P = \sum_{i=1}^{n} h_i \sigma_i^z + \sum_{i<j} J_{ij}\sigma_i^z\sigma_j^z,
\end{equation}
where $h_i$ and $J_{ij}$ are the linear and quadratic Ising coefficients. 

The qubit register is initialized in the equal superposition state $\ket{+}^{\otimes n}$, where $\ket{+}=(\ket{0}+\ket{1})/\sqrt{2}$. For each layer $j$, the problem and driver unitaries are given by $e^{-i\gamma_j H_P}$ and $e^{i\beta_j H_D}$, respectively, with $\gamma_j$ and $\beta_j$ the circuit parameters. After applying $p$ alternating layers of these unitaries, the resulting parametrized state is
\begin{equation}\label{eq:QAOAFinalState}
\ket{\boldsymbol{\gamma},\boldsymbol{\beta}}
=
\prod_{j=1}^{p}
e^{i\beta_j H_D}
e^{-i\gamma_j H_P}
\ket{+}^{\otimes n},
\end{equation}
where the product is ordered such that the layer with $j=1$ is the rightmost term. 

QAOA is usually viewed as a variational algorithm in which the parameters $\boldsymbol{\gamma} = \{\gamma_1, \ldots, \gamma_p\}$ and $\boldsymbol{\beta} = \{\beta_1, \ldots, \beta_p\}$ are optimized to minimize the expectation value $\bra{\boldsymbol{\gamma},\boldsymbol{\beta}} H_P \ket{\boldsymbol{\gamma},\boldsymbol{\beta}}$. Alternatively, the parameters can be fixed according to a predetermined schedule. Specifically, in the linear ramp QAOA (LR-QAOA) protocol \cite{MontlinearrampQAOA2025}, the parameters are set to
\begin{equation}\label{eq:QAOAParams}
\gamma_j = \frac{j}{p}\Delta_\gamma,
\qquad
\beta_j = \left(1-\frac{j-1}{p}\right)\Delta_\beta,
\end{equation}
for $j=1,\ldots,p$, and we fix the ramp amplitudes to be $\Delta_\gamma=\Delta_\beta=1$.

LR-QAOA can also be seen as a discretized version of quantum annealing with a linear annealing schedule. As such, this approach has been studied in the literature under different names as well, including Approximate Quantum Annealing (AQA) \cite{WillGPUacceleratedsimulations2022} and Trotterized Quantum Annealing (TQA) \cite{SackQuantumannealing2021}. Here, we choose LR-QAOA instead of the variational approach to QAOA to compare the different QUBO formulations of the Knapsack problem. This allows us to attribute differences in performance to the QUBO formulations themselves, rather than to differences arising from parameter optimization. 

\subsection{D-Wave}\label{subsec:DWave}
D-Wave Quantum Processing Units (QPUs) provide a practical implementation of quantum annealing, which can be described by the time-dependent Hamiltonian
\begin{equation}
    H(s) = A(s) H_{\mathrm{D}} + B(s) H_{\mathrm{P}},
    \label{eq:dwave_hamiltonian}
\end{equation}
where $s\in[0,1]$ is the annealing parameter and $A(s)$ and $B(s)$ form the so-called annealing schedule, with $A(s)$ decreasing and $B(s)$ increasing with $s$, such that $A(0)\gg B(0)$ and $A(1)\ll B(1)$. The driver Hamiltonian is, as for QAOA, the transverse-field Hamiltonian $H_{\mathrm{D}}=-\sum_i \sigma_i^x$, and the problem Hamiltonian is again given by \cref{eq:ProblemHam}.

If the problem graph of an optimization problem does not match the hardware graph of the QPU, a minor embedding is required to map the problem onto the hardware \cite{ChoiMinorembeddingadiabatic2008}. As long as the size of the optimization problem is small enough, several copies of the problem can be embedded onto the hardware simultaneously, allowing for parallel execution. The D-Wave systems provide several additional controls over the annealing evolution, but other than this parallelization, we use the default settings for all experiments in this paper and set the number of samples to around 10000 per run. Like for LR-QAOA, any differences in performance can then be attributed to the QUBO formulations themselves rather than to differences in these settings.

\subsection{Problem instance generation}\label{subsec:ProblemInstances}
We generate instances of \KP by drawing the item weights $w_i$ and values $v_i$ uniformly at random from the set of integers between 1 and 10. The knapsack capacity $W$ is set to 70\% of the total weight of all items, rounded down to the nearest integer. The problem generation procedure follows that in the Knapsack repository of the JUNIQ Benchmark Suite \cite{JBS}. For each problem size $n$ in the range $n \in [4, 20]$, we generate 50 random instances of \KP.

These instances are converted into \QKP instances using the binary encoding for the slack variable $s$, as described in \cref{subsec:SlackEncoding}. The range of the slack variable is either the standard range $s \in \{0,1,\dots,W\}$ (see \cref{subsec:KnapsackQUBO}) or the reduced range $s \in \{0,1,\dots,w_{max}-1\}$ (see \cref{subsec:ReduceSlack}). The penalty constant for the \QKP instances is set to either $(1+\alpha) v_{max}$ (\cref{eq:VmaxBound}) or $(1+\alpha) v_s$ (\cref{eq:VsBound}), with $\alpha = 0.001$. After conversion of the \QKP instances into Ising form through the substitution $x_i \to (1 - \sigma_i)/2$, we normalize the resulting Ising problems, by dividing all coefficients $h_i$ and $J_{ij}$ by the largest absolute coefficient $\max\left\{ \max_i |h_i|,\ \max_{i<j} |J_{ij}| \right\}$. 

The slackless method is implemented according to the description in \cref{subsec:Implementation}. For this method too, all penalty constants are multiplied by the factor $(1+\alpha)$, with $\alpha = 0.001$. The resulting Ising problems are normalized in the same way as for the \QKP instances.

\section{Results and discussion}\label{sec:Results}

This section is structured according to the three proposed approaches to improve the standard \QKP formulation for the Knapsack problem, which were presented in \cref{sec:ImprovedMult,sec:ReducedSlack,sec:SlacklessForm}.  

\subsection{Smaller penalty constant}\label{subsec:PenaltyConstantSmaller}

\begin{figure*}
    \centering
    \includegraphics[width=\textwidth]{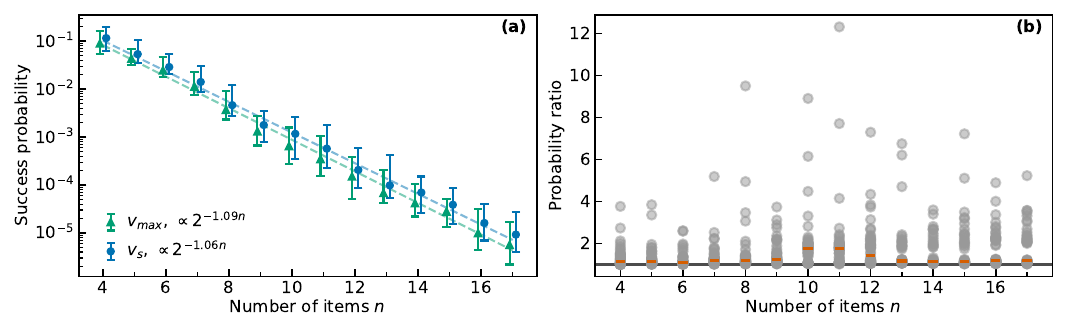}
    \caption{Comparison of the performance of LR-QAOA with $p=10$ for the two penalty constants $\lambda_0$ considered for \QKP: $v_{max}$ from \cref{eq:VmaxBound} and $v_s$ from \cref{eq:VsBound}. (a) Dependence of the probability of sampling an optimal solution to \KP on problem size. The markers indicate the median over the problem instances generated for each $n$, and error bars indicate the 10th to 90th percentiles. Exponential fits are performed on the median success probabilities. Markers are horizontally offset to improve visual clarity. (b) Success probability for the $v_s$ formulation divided by that for $v_{max}$ for each problem instance. Orange horizontal bars indicate the median ratio and the black horizontal line indicates a ratio of 1.
    }
    \label{fig:PenaltyConstantQAOA}
\end{figure*}

We compare the \QKP formulation with the smaller penalty constant $v_s$, see \cref{eq:VsBound} in \cref{sec:ImprovedMult}, against the standard penalty constant $v_{max}$, see \cref{eq:VmaxBound} in \cref{subsec:PenaltyConstant}. For both formulations, the slack range is restricted using the technique shown in \cref{subsec:ReduceSlack}, to reduce the required number of slack variables.

\subsubsection{LR-QAOA}\label{subsubsec:PenaltyConstantQAOA}

We first investigate the effect of the two penalty constants on the performance of LR-QAOA with $p=10$. The median success probability over the problem instances generated for each $n$ decreases exponentially with problem size, as shown in \cref{fig:PenaltyConstantQAOA}(a). In fact, for both penalty constants, the success probability scales with $n$ approximately as it does for randomly guessing the solution, namely as $2^{-n}$. The smaller penalty constant thus does not seem to influence this scaling, although it seems to consistently improve the success probability by a constant factor. 

In \cref{fig:PenaltyConstantQAOA}(b), the ratio between the success probability for the smaller constant $v_s$ and that for $v_{max}$ is shown for each problem instance in our set. This figure shows that the smaller penalty constant $v_s$ indeed increases the success probability by a factor that seems independent of the problem size $n$. Interestingly, there is not a single problem in our problem set for which the success probability is not improved when using $v_s$ instead of $v_{max}$.

Since the LR-QAOA protocol does not involve any classical optimization procedure, the cost Hamiltonian is the only component of the algorithm that is altered when changing the penalty constant. The cost Hamiltonian that is constructed with the smaller penalty constant $v_s$ thus appears to be consistently more favorable for LR-QAOA than the one constructed with $v_{max}$, at least for the set of problems considered here. 

Note that, since we randomly generate problem instances, some instances have multiple optimal solutions. For those problems, we report as the success probability the sum of the probabilities of sampling each optimal solution. This variation in the number of optimal solutions across instances may contribute to the spread in success probabilities for each $n$, as observed, e.g., in \cref{fig:PenaltyConstantQAOA}(a).

Moreover, each optimal solution to the \KP problem gives rise to multiple solutions to the \QKP problem, all with different slack variables. That is, these solutions have the same $\mathbf{x}$-component that is optimal to \KP, while their slack variables result in different values of the penalty term in \cref{eq:KP_penalty}. We also count all such solutions as optimal when calculating the success probability, since all of them still solve the original \KP problem. The two conventions described above for defining the success probability apply to all results reported in this paper.

\subsubsection{D-Wave}\label{subsubsec:PenaltyConstantDWave}

\begin{figure*}
    \centering
    \includegraphics[width=\textwidth]{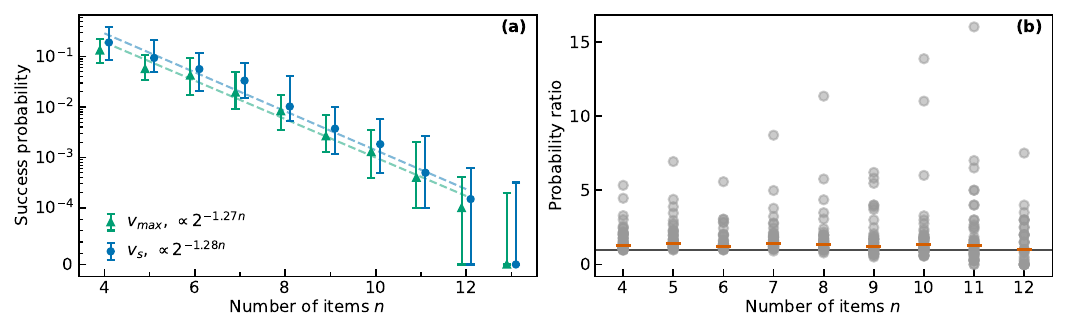}
    \caption{Same as \cref{fig:PenaltyConstantQAOA}, but for the D-Wave device. Note that the vertical axis in (a) uses a linear scale below $10^{-4}$ and a logarithmic scale above. 
    }
    \label{fig:PenaltyConstantDWave}
\end{figure*}

On the D-Wave quantum annealer too, the median success probability decreases exponentially with problem size for both penalty constants, as shown in \cref{fig:PenaltyConstantDWave}(a). The exponential decay is steeper than for LR-QAOA. 
The need of embedding the problems onto the hardware graph of the D-Wave device \cite{ChoiMinorembeddingadiabatic2008} may contribute to this, as it introduces additional qubits, effectively increasing the problem size. The scaling of the median success probability with $n$ is again similar for the two penalty constants, with the smaller penalty constant $v_s$ consistently yielding the higher median success probability.

When considering the ratio between the success probabilities for individual problem instances, as shown in \cref{fig:PenaltyConstantDWave}(b), we indeed find that using the smaller constant $v_s$ improves the success probability for most problem instances. Problem instances with $n=13$ were excluded since more than half yield an undefined ratio due to a zero denominator, and this occurs for $24$ problems with $n < 13$ too. In contrast to the LR-QAOA results, the success probability is lower when using $v_s$ than $v_{max}$ for $64$ of the $426$ problem instances plotted in \cref{fig:PenaltyConstantDWave}(b). However, the majority of the problem instances for which this occurs have low success probabilities for both penalty constants, resulting in a large uncertainty in the ratio. The median ratio remains greater than or equal to $1$ for all problem sizes (including for $n=13$).

\subsection{Reduced slack range}\label{subsec:ReducedSlackRange}

\begin{figure*}
    \centering
    \includegraphics[width=\textwidth]{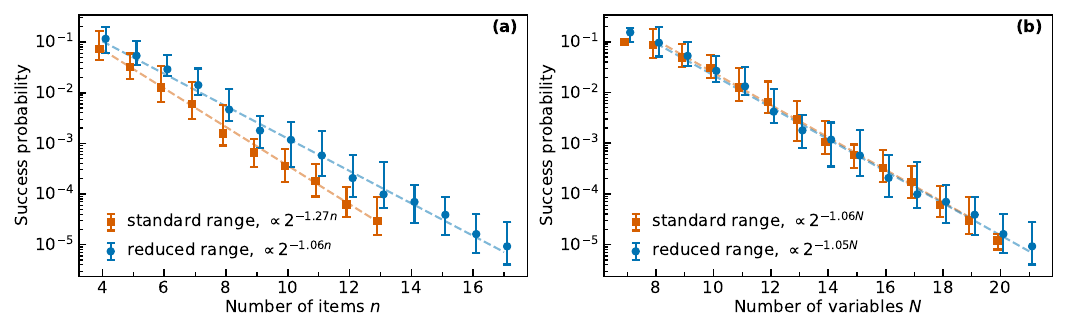}
    \caption{Dependence of the probability of sampling an optimal solution to \KP on (a) problem size $n$ and (b) the number of variables $N$, for the two slack ranges considered for \QKP: $s \in \{0,1,\dots,W\}$ (standard) and $s \in \{0,1,\dots,w_{max}-1\}$ (reduced). Results obtained using simulations of LR-QAOA with $p=10$. The markers indicate the median over the problem set for each $n$, and error bars indicate the 10th to 90th percentiles. Exponential fits are performed on the median success probabilities, excluding problem sizes $N$ with fewer than $10$ problem instances. Markers are horizontally offset to improve visual clarity.}
    \label{fig:ReducedSlackRangeQAOA}
\end{figure*}

We compare the \QKP formulation using the standard slack range $s \in \{0,1,\dots,W\}$ (see \cref{subsec:KnapsackQUBO}) against the formulation using the reduced slack range $s \in \{0,1,\dots,w_{max}-1\}$ (see \cref{subsec:ReducedSlackRange}). The smaller penalty constant $v_s$, see \cref{eq:VsBound} in \cref{sec:ImprovedMult}, is used for both formulations.

\subsubsection{LR-QAOA}\label{subsubsec:ReducedSlackRangeQAOA}

The performance of LR-QAOA with $p=10$ for the two slack variable ranges is shown in \cref{fig:ReducedSlackRangeQAOA}(a). The reduced slack range yields higher median success probabilities than the standard range, where the median is taken over the set of problems for each $n$. For only $11$ of the $500$ problem instances that were solved with both slack ranges (those problems with $n \leq 13$), the success probability is lower for the reduced range than for the standard range. Moreover, the decay in success probability with $n$ is slower for the reduced range, as evidenced by the smaller exponent of the exponential fit. The fact that the \QKP formulation with the reduced slack range assigns a penalty to feasible solutions with total weight below $W - w_{max}$ (see \cref{fig:PenaltySketch}(b)) thus does not appear to be detrimental to performance. 

When the success probability is instead plotted against the number of variables $N$, which consists of the $n$ item variables together with the slack variables, as shown in \cref{fig:ReducedSlackRangeQAOA}(b), the median success probabilities for the two slack ranges are similar. 
This suggests that the improved performance of the reduced slack range observed in \cref{fig:ReducedSlackRangeQAOA}(a) is mainly due to the reduction in the number of slack variables (see \cref{subsec:ReduceSlack}), as its advantage disappears when the two formulations are compared at the same number of variables. For a given problem of size $n$, the reduced slack range requires a smaller number of variables $N$, which, according to \cref{fig:ReducedSlackRangeQAOA}(b), is expected to result in a higher success probability. 
The smaller number of slack variables for the reduced range also allows larger problem sizes to be solved with comparable computational resources (the problems with $n \geq 14$ in \cref{fig:ReducedSlackRangeQAOA}(a)). 

The number of slack variables required for the \QKP formulation with the reduced slack range is determined by the maximum item weight $w_{max}$ (see \cref{subsec:ReduceSlack}), which, in our case, is bounded by the maximum value of the uniform distribution from which the item weights are drawn (see \cref{subsec:ProblemInstances}). Consequently, there is an upper bound on the number of slack variables required, which is independent of $n$. Out of the $700$ problems considered here, $657$ require the maximum of $4$ slack variables. The exponential decay in success probability is therefore similar in \cref{fig:ReducedSlackRangeQAOA}(a) and \cref{fig:ReducedSlackRangeQAOA}(b); most data points are simply horizontally offset by the $4$ slack variables.

In contrast, the number of slack variables required for the standard slack range is determined by the knapsack capacity $W$ (see \cref{subsec:SlackEncoding}). Since $W$ is chosen as $70\%$ of the total weight of all items, it increases with the problem size $n$, and therefore so does the number of slack variables, which scales approximately as $\log(W)$. Given that the success probability is similar for a given $N$ for both slack ranges (\cref{fig:ReducedSlackRangeQAOA}(b)), the increase in the number of variables $N$ with $n$ for the standard range, in contrast to $N$ remaining independent on $n$ for the reduced range, can explain the faster decrease in success probability with $n$ observed in \cref{fig:ReducedSlackRangeQAOA}(a).

\subsubsection{D-Wave}\label{subsubsec:RecucedSlackRangeDWave}

\begin{figure*}
    \centering
    \includegraphics[width=\textwidth]{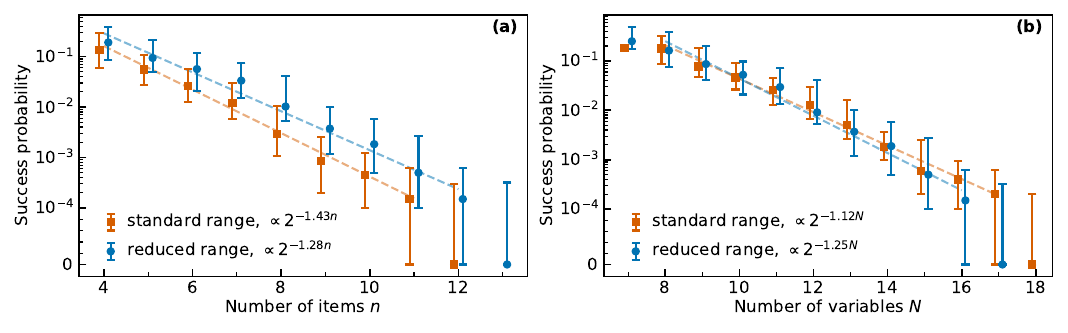}
    \caption{Same as \cref{fig:ReducedSlackRangeQAOA}, but for the D-Wave device. Note that the vertical axis in (a) uses a linear scale below $10^{-4}$ and a logarithmic scale above.
    }
    \label{fig:ReducedSlackRangeDWave}
\end{figure*}

The performance of the two slack variable ranges on the D-Wave device is compared in \cref{fig:ReducedSlackRangeDWave}(a). The reduced slack range yields higher median success probabilities than the standard range for each $n$. Only for $23$ of the $450$ problem instances that were solved with both ranges (those problems with $n \leq 12$,) the success probability is lower for the reduced range than for the standard range. The exponential decay of the median success probabilities is more rapid for the standard range than for the reduced range, consistent with the LR-QAOA results.

When plotted against the number of variables $N$, the median success probabilities for the two slack ranges are similar, as shown in \cref{fig:ReducedSlackRangeDWave}(b), as was observed for LR-QAOA. This suggests that, on the D-Wave device as well, the main contribution to the improved performance of the reduced slack range is the reduction in the number of slack variables. And the difference in the exponential decay rates between the two slack ranges in \cref{fig:ReducedSlackRangeDWave}(a) can, as before, be attributed to the different dependence of the number of slack variables, and thus $N$, on the problem size $n$ for the two slack ranges.

\subsection{Slackless method}\label{subsec:SlacklessMethodResults}

\begin{figure*}
    \centering
    \includegraphics[width=\textwidth]{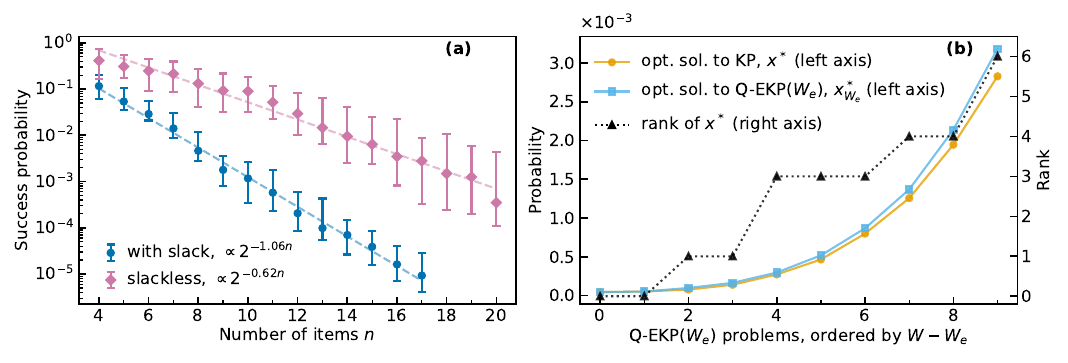}
    \caption{(a) Dependence of the probability of sampling an optimal solution to \KP on problem size $n$, for the slackless method and the \QKP formulation (with slack) with the smaller penalty constant $v_s$ (\cref{eq:VsBound}) and the reduced slack range $s \in \{0,1,\dots,w_{max}-1\}$, obtained from simulations of LR-QAOA with $p=10$. For the slackless method, the average success probability over all $\QEKP[W_e] \in \mathcal{Q}$ is taken for each problem instance. The markers indicate the median over the problem set for each $n$, and error bars indicate the 10th to 90th percentiles. Exponential fits are performed on the median success probabilities. (b) Probability of sampling $\mathbf{x^{*}}$, the optimal solution to \KP (the same for all instances) and $\mathbf{x^{*}_{W_e}}$, the optimal solution to each $\QEKP[W_e] \in \mathcal{Q}$, shown on the left axis. The rank of $\mathbf{x}^{*}$, i.e., its index among all solutions to $\QEKP[W_E]$ ordered by decreasing objective value (rank $0$ corresponds to the highest objective value), is shown on the right axis. Both quantities are shown for each problem $\QEKP[W_e] \in \mathcal{Q}$, for an example \KP instance with $n=16$ that has a single optimal solution.}
    \label{fig:SlacklessMethodsQAOA}
\end{figure*}

We compare the \QKP formulation with the smaller penalty constant $v_s$ (\cref{eq:VsBound}) and the reduced slack range $s \in \{0,1,\dots,w_{max}-1\}$ (see \cref{subsec:ReduceSlack}) against the slackless method, which was presented in \cref{sec:SlacklessForm}.

\subsubsection{LR-QAOA}\label{subsubsec:SlacklessMethodQAOA}
The performance of LR-QAOA with $p=10$, in terms of success probability, is shown in \cref{fig:SlacklessMethodsQAOA}(a) for both methods. Since the slackless method requires solving all $w_{max}$ problems in the set $\mathcal{Q}$ (see \cref{subsec:Implementation}), no single success probability is associated with this method. Instead, we report the average success probability over the set $\mathcal{Q}$ for this method. The slackless method then yields higher success probabilities than the \QKP formulation for all problem instances, with the difference in the median probabilities reaching several orders of magnitude for larger $n$. Moreover, the exponential decay of the median success probability is significantly slower for the slackless method. The spread of success probabilities across problem instances, on the other hand, is larger for the slackless method than for the \QKP formulation. Overall, the slackless method achieves higher success probabilities, even though it has the same worst-case asymptotic runtime as the \QKP formulation for a solver with exponential runtime (see \cref{subsec:CompAndDisc}).

The average success probability over the set $\mathcal{Q}$ can be interpreted as the success probability of a randomized procedure for the slackless method. In each iteration, this procedure selects a problem $\QEKP[W_E] \in \mathcal{Q}$ uniformly at random and performs a single run of LR-QAOA, or another solver, on the selected problem. The expected success probability for each iteration is then exactly the average of the success probabilities for the problems in $\mathcal{Q}$.

We remark that there is the potential to design more optimized procedures for the slackless method. For example, a small number of runs could first be performed on each problem in $\mathcal{Q}$. A metric such as the average objective value over these runs could then be used to identify the problems that are more likely to yield solutions that are optimal to $\KP$, and later runs could focus on those problems. In this way, the success probability of the slackless method could effectively be increased beyond the average success probability over the set $\mathcal{Q}$. The design of such procedures is left for future work. The results in \cref{fig:SlacklessMethodsQAOA}(a) nonetheless show that even a simple randomized procedure can achieve a higher success probability than the $\QKP$ formulation.

For an example \KP instance, \cref{fig:SlacklessMethodsQAOA}(b) shows the performance of each individual problem \QEKP[W_e] in the set $\mathcal{Q}$, constructed using the slackless method. For this particular instance, there is a single optimal solution (to \KP) $\mathbf{x}^{*}$ with total weight $W^{*} = W$, so only the corresponding problem $\QEKP[W]$ is guaranteed to be a \textit{valid} reformulation (see \cref{subsec:Implementation}), i.e., its optimal solution $\mathbf{x}^{*}_W$ is equal to $\mathbf{x}^{*}$. While one other problem $\QEKP[W-1]$ also has $\mathbf{x}^{*}$ as its optimal solution, for the other problems $\mathbf{x}^{*}$ has a lower objective value than $\mathbf{x}^{*}_{W_e}$, and thus has a higher rank in the objective-value ordering. For those problems, the probability of sampling $\mathbf{x}^{*}$ is indeed lower than that of sampling $\mathbf{x}^{*}_{W_e}$. 

However, as $W_e$ decreases, the corresponding problems \QEKP[W_e] have a higher probability to sample $\mathbf{x}^{*}$, even though the rank of $\mathbf{x}^{*}$ increases at the same time. Thus, the problems \QEKP[W_e] whose optimal solutions are not optimal to \KP, surprisingly yield higher probabilities of sampling $\mathbf{x}^{*}$ than the problems whose optimal solutions are optimal to \KP. At the same time, the probability to sample $\mathbf{x}^{*}_{W_e}$ increases with decreasing $W_e$ too, suggesting that, for the corresponding problems \QEKP[W_e], the sampling distributions increasingly shift towards solutions with high objective values.

We observe the pattern of increasing success probability (the probability of sampling $\mathbf{x}^{*}$) for problems $\QEKP[W_e] \in \mathcal{Q}$ with decreasing $W_e$ for many problem instances, not just for this example instance. This pattern is more evident for larger problem sizes; in \cref{app:SlacklessMethodPerformance}, we show that it occurs for all instances with $n=16$. Thus, even though the optimal solutions to the problems $\QEKP[W_e] \in \mathcal{Q}$ with small $W_e$ may not be optimal to \KP, the presence of these problems in $\mathcal{Q}$ is beneficial for the slackless method, as it increases the average success probability.

One possible explanation for this pattern of higher success probabilities at smaller $W_e$ is that the penalty constants for the problems in $\mathcal{Q}$ also decrease as $W_e$ becomes smaller (see \cref{eq:Lambda1BoundValues,eq:Lambda1BoundWeightFinal} and \cref{subsec:Implementation}). This would be consistent with previous observations on the importance of penalty constants \cite{RochCrossEntropy2020,RochCrossEntropy2021,RochEffectPenalty2023,VillAnalyzingbehaviour2022,GuneQUBOFormulations2025,AlesAlleviatingquantum2025}, although it is remarkable that the value of the penalty constant appears to have a stronger effect on the success probability than whether the solutions $\mathbf{x}^{*}_{W_e}$ to the problems $\QEKP[W_e] \in \mathcal{Q}$ are actually optimal to \KP.

Note that in simulations of LR-QAOA, one would expect the probability of sampling $\mathbf{x}^{*}$ to go to $1$ as $p$ goes to infinity \cite{FarhQuantumApproximate2014,ZhouQuantumApproximate2020,WillBenchmarkingquantum2020} for any problem $\QEKP[W_e] \in \mathcal{Q}$ that has $\mathbf{x}^{*}$ as its optimal solution, and similarly the probability of sampling $\mathbf{x}^{*}_{W_e}$ to go to $1$ for all other problems in $\mathcal{Q}$. In this limit, the problems \QEKP[W_e] whose optimal solutions are not optimal to \KP would therefore not contribute to the average success probability of the slackless method anymore. However, the depth of the circuits that can be faithfully executed on present-day quantum devices is limited, and consequently the number of layers $p$ in LR-QAOA too. For the value $p=10$ considered here, the success probabilities are still far from $1$ for all but the smallest problem sizes, indicating that we have not yet reached the large-$p$ regime. Our results therefore reflect some intermediate-$p$ regime relevant to current quantum hardware, and in this regime the slackless method outperforms the \QKP formulation.

\subsubsection{D-Wave}\label{subsubsec:SlacklessMethodDWave}

\begin{figure*}
    \centering
    \includegraphics[width=\textwidth]{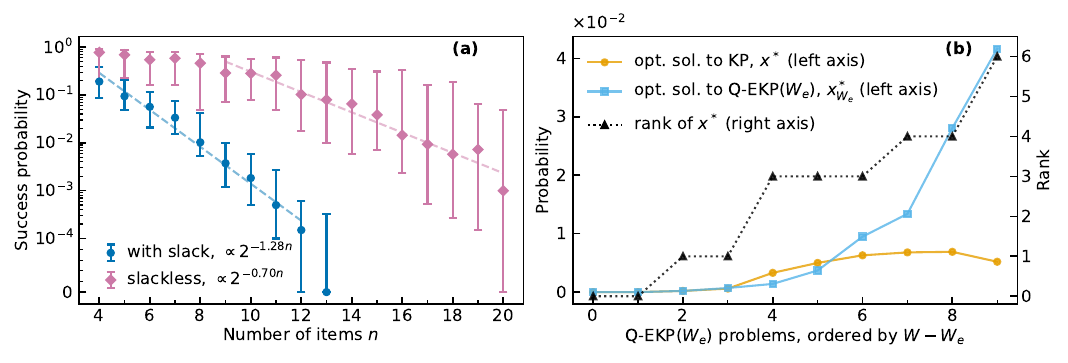}
    \caption{Same as \cref{fig:SlacklessMethodsQAOA}, but for the D-Wave device. Note that the vertical axis in (a) uses a linear scale below $10^{-4}$ and a logarithmic scale above. The same example \KP instance as in \cref{fig:SlacklessMethodsQAOA}(b) is used for (b).
    }
    \label{fig:SlacklessMethodsDWave}
\end{figure*}

In \cref{fig:SlacklessMethodsDWave}(a), the performance of the slackless method is compared against that for the \QKP formulation on the D-Wave device. For the slackless method, the average success probability over the set $\mathcal{Q}$ is reported. As for LR-QAOA, the success probability for the slackless method is higher than that for the \QKP formulation for all problem instances, except for a single problem with $n=4$, and the exponential decay of the median success probability is significantly slower for the slackless method. This also means that the slackless method can successfully solve larger problems than the \QKP formulation when using the same number of samples. The variation across problem instances is larger for the slackless method than for the \QKP formulation, and is also larger than for LR-QAOA.

In \cref{fig:SlacklessMethodsDWave}(b), the performance of each individual problem $\QEKP[W_e] \in \mathcal{Q}$ is shown for the same individual \KP instance as in \cref{fig:SlacklessMethodsQAOA}(b). As for LR-QAOA, the probability to sample $\mathbf{x}^{*}$ is higher for the problems $\QEKP[W_e]$ with smaller $W_e$, whose optimal solutions $\mathbf{x}^{*}_{W_e}$ are not optimal to \KP, than for the two problems with highest $W_e$, whose optimal solutions are optimal to \KP. 
However, for this problem instance, the probability of sampling $\mathbf{x}^{*}$ does not follow the probability of sampling $\mathbf{x}^{*}_{W_e}$ closely, in contrast to the LR-QAOA results.

In \cref{app:SlacklessMethodPerformance} we show that the pattern of higher success probabilities for problems $\QEKP[W_e] \in \mathcal{Q}$ with smaller $W_e$ is less consistently observed across problem instances for the D-Wave device than for LR-QAOA. For the instances for which this pattern is observed, the value of the penalty constant, which decreases with decreasing $W_e$, may again have a greater influence on the success probability than whether the optimal solutions to the problems $\QEKP[W_e] \in \mathcal{Q}$ are optimal to \KP. 

\section{Conclusion}\label{sec:Conclusion}

We have proposed three approaches for improving the standard QUBO formulation of the Knapsack problem: (i) a smaller penalty constant, (ii) a reduction of the number of slack variables and (iii) a slackless method, which solves a set of QUBO problems without slack variables. The smaller penalty constant is derived using a general approach that we also apply to derive the penalty constants for the slackless method. The reduction in the number of slack variables is possible because of a property of the optimal solution to the Knapsack problem. 

The slackless method constructs a set of equality-constrained Knapsack problems, each formulated as a QUBO problem without additional slack variables. The penalty constants are chosen such that at least one of these problems shares an optimal solution with the original Knapsack problem instance. However, since it is not known in advance for which problem(s) this is the case, the slackless method requires solving all problems in the set. 

We have shown that all three approaches lead to an improvement in the median success probability over our problem set when using either the LR-QAOA algorithm (with $p=10$) or the D-Wave device. The smaller penalty constant increases the success probability by a factor that appears to be independent of the problem size. The improvement in success probability from reducing the number of slack variables, on the other hand, increases with problem size, since the number of slack variables saved also increases with problem size. 

The slackless method achieves the largest improvement out of the three approaches. The exponential scaling of the success probability with problem size is significantly more favourable for the slackless method than for the standard QUBO formulation after applying the other two improvements. Even at the relatively small problem sizes considered here, success probabilities are several orders of magnitude higher.

For many problem instances, we find that, among the problems in the set constructed by the slackless method, those with smaller penalty constants contribute most to the success probability of the method, even when their optimal solutions are not optimal to the original Knapsack problem.
While this observation would not be expected in a regime where the probability to find the optimal solution to each problem is close to $1$, it appears to give the slackless method an advantage over the standard QUBO formulation both on the D-Wave device and for LR-QAOA simulations in the intermediate-$p$ regime, which corresponds to what is currently feasible on quantum hardware.

The results presented here for the slackless method implicitly correspond to a simple procedure in which each problem in the constructed set is selected with equal probability before each run of the chosen algorithm. The slackless method thus has considerable potential for further development through more sophisticated procedures for allocating the available computational resources. For example, one could first identify the problems in the set that are more likely to produce optimal solutions to the original Knapsack problem and assign more resources to those problems. We leave the development of such procedures for future work.

Collectively, the results for the three proposed improvements demonstrate that the formulation of the QUBO problem can have a significant effect on the performance of quantum solvers. In particular, the performance of the slackless method demonstrates that, even though it is not expected to outperform the standard formulation when both are solved with an exponential-runtime solver, it can nevertheless lead to substantial improvements for quantum optimization on current quantum hardware.

\section{Acknowledgements}
The authors acknowledge support from the project Entwicklungspartnerschaft Ionenfallen-Quantencomputer in NRW (EPIQ), funded by the Ministerium für Kultur und Wissenschaft des Landes Nordrhein-Westfalen (MKW NRW). P.v.d.H. and J.A.M. acknowledge support from the Federal Ministry of Research, Technology and Space (BMFTR) within the framework programme “Quantum technologies - from basic research to market” through the project QSolid (Grant No. 13N16149). The authors gratefully acknowledge the Jülich Supercomputing Centre (https://www.fz-juelich.de/ias/jsc) for funding this project by providing computing time on the D-Wave Advantage™ System JUPSI through the Jülich UNified Infrastructure for Quantum computing (JUNIQ). 
D.W. acknowledges support from the project Jülich UNified Infrastructure for Quantum Computing (JUNIQ) that has received funding from the German Federal Ministry of Education and Research (BMBF) and the Ministry of Culture and Science of the State of North Rhine-Westphalia.
D.W. acknowledges support from the project QEC4QEA that has received funding from the European High Performance Computing Joint Undertaking (EuroHPC JU) under Grant Agreement No 101194322, and
from the project “Quantum-based Energy Grids (QuGrids),” funded under the programme “Profilbildung 2022,” an initiative of the Ministry of Culture and Science of the State of North Rhine-Westphalia.
D.W. acknowledges support from Q-NEKO through the EuroHPC Joint Undertaking under Grant Agreement No. 101241875 and its members, including top-up funding by the Government of Japan, and was co-funded by the European Union.

\appendix

\section{Greedy bound proof}\label[appendix]{app:GreedyBoundProof}
In this appendix we show that the ratio-based greedy algorithm can provide a tighter lower bound $W_R$ of the optimal solution's weight $W^{*}$ than the bound $W - w_{max}$ introduced in \cref{subsec:ReduceSlack}. This algorithm \cite{KellKnapsackProblems2004} ranks the items according to their value to weight ratio $v_i / w_i$ (also called the efficiency, with $v_i/w_i > 0$ due to our choice of $v_i$ and $w_i$), goes through the items in decreasing order of this ratio, including all items in the knapsack until it encounters an item that would exceed the knapsack capacity, then terminates. It returns the set of items $\mathcal{G}_R$, which corresponds to the greedy solution $\mathbf{x}^R$ ($x^R_i = 1 \; \forall \; i \in \mathcal{G}_R$). 

We show below that $\mathbf{x}^R$ is optimal for the knapsack problem with capacity $W_{R} = \sum_i w_i x_i^R$ instead of $W$ 
(since $\mathbf{x}^R$ cannot be improved by exchanging any of its items with a less efficient item). 
Only solutions whose weight exceeds $W_R$ can then have a higher objective value than $\mathbf{x}^R$. Therefore, $W_R$ provides an improved lower bound for the total weight of the optimal solution: $W^{*} \geq W_R > W - w_{\max}$.
Using this new bound, the number of slack variables can be reduced in the same way as shown in \cref{subsec:ReduceSlack}. Namely, the slack variable $s$ is taken to lie in $\{0, 1, \dots, W - W_R\}$, requiring $\lfloor \log(W - W_R) \rfloor + 1$ slack variables. 

We now show that $\mathbf{x}^R$ is optimal for the knapsack problem with capacity $W_R$, by showing that any (other) feasible solution $\hat{\mathbf{x}}$ to the problem with capacity $W_R$ (satisfying $\sum_i w_i \hat{x}_i \leq W_R$) cannot have a greater objective value than that of $\mathbf{x}^{R}$.

First, we order the items according to their value-to-weight ratio $v_i / w_i$:
\begin{equation}\label{eq:ItemsOrdering}
    \frac{v_1}{w_1} \geq \frac{v_2}{w_2} \geq \ldots \geq \frac{v_n}{w_n}.
\end{equation}
As mentioned, the greedy algorithm selects items in decreasing order of this ratio and terminates once adding the next item $r$ would exceed the capacity. We label the final item that is included $r-1$. The value-to-weight ratio of the items in $\mathcal{G}_R$ is then at least $v_{r-1} / w_{r-1}$, while items in the complementary set $\bar{\mathcal{G}}_R$ have a maximal ratio of $v_r / w_r$. Any feasible solution $\hat{\mathbf{x}}$ can be constructed from $\mathbf{x}^R$ by removing a set of items $\mathcal{S}_{rem} \subseteq \mathcal{G}_R$ from it and replacing it with a set of items $\mathcal{S}_{add} \subseteq \bar{\mathcal{G}}_R$. The value-to-weight ratio bounds for the sets $\mathcal{G}_R$ and $\bar{\mathcal{G}}_R$ naturally also apply to their respective subsets $\mathcal{S}_{rem}$ and $\mathcal{S}_{add}$. Thus, the objective value \eqref{eq:KP-obj} of these sets can be bounded as follows:

\begin{equation}\label{eq:GreedySetInequality}
    \sum_{i \in S_{rem}} v_i = \sum_{i \in S_{rem}} \frac{v_i}{w_i} w_i \geq \frac{v_{r-1}}{w_{r-1}} \sum_{i \in S_{rem}} w_i,
\end{equation}
and
\begin{equation}\label{eq:NotGreedySetInequality}
    \sum_{i \in S_{add}} v_i = \sum_{i \in S_{add}} \frac{v_i}{w_i} w_i \leq \frac{v_r}{w_r} \sum_{i \in S_{add}} w_i.
\end{equation}

The total weight of $\mathbf{x}^R$ equals the capacity $W_R$, so the total weight of $\hat{\mathbf{x}}$ is only feasible when the total weight is not increased: $\sum_{i \in S_{add}} w_i \leq \sum_{i \in S_{rem}} w_i$. Combining this with \cref{eq:GreedySetInequality} and \cref{eq:NotGreedySetInequality}, we obtain: 
\begin{multline}\label{eq:ProofGreedyIsOptimal}
    \sum_{i \in \mathcal{S}_{add}} v_i 
    \leq \frac{v_r}{w_r} \sum_{i \in \mathcal{S}_{add}} w_i 
    \leq \frac{v_r}{w_r} \sum_{i \in \mathcal{S}_{rem}} w_i \\
    \leq \frac{v_{r-1}}{w_{r-1}} \sum_{i \in \mathcal{S}_{rem}} w_i 
    \leq \sum_{i \in \mathcal{S}_{rem}} v_i.
\end{multline}
Therefore, the objective value of any feasible solution $\hat{\mathbf{x}}$ can be no greater than that of $\mathbf{x}^R$;
$\mathbf{x}^R$ must be an optimal solution to the Knapsack problem with capacity $W_R$ (exact equality in \cref{eq:ProofGreedyIsOptimal} is only attained when all items in both $\mathcal{S}_{rem}$ and $ \mathcal{S}_{add}$ have the same ratio $v_r / w_r = v_{r-1} / w_{r-1}$ and the total weight of the sets is equal ($\sum_{i \in S_{rem}} = \sum_{i \in S_{add}}$), in which case $\hat{\mathbf{x}}$ and $\mathbf{x}^R$ are both optimal solutions).

\section{Performance of the slackless method across instances}\label[appendix]{app:SlacklessMethodPerformance}

\begin{figure}
    \centering
    \includegraphics[width=\columnwidth]{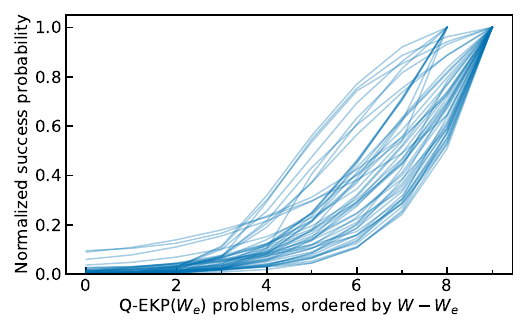}
    \caption{Probability of sampling the optimal solution $\mathbf{x}^{*}$ to $\KP$ using LR-QAOA with $p=10$, shown for all 50 problem instances generated for $n=16$. For each instance, the probability is shown for each $\QEKP[W_e] \in \mathcal{Q}$, with $\mathcal{Q}$ generated according to the slackless method (see \cref{subsec:Implementation}). The probabilities are normalized independently for each problem instance by the maximum probability observed across the set $\mathcal{Q}$.}
    \label{fig:NormalizedSuccessProbabilitiesQAOA}
\end{figure}

\begin{figure}
    \centering
    \includegraphics[width=\columnwidth]{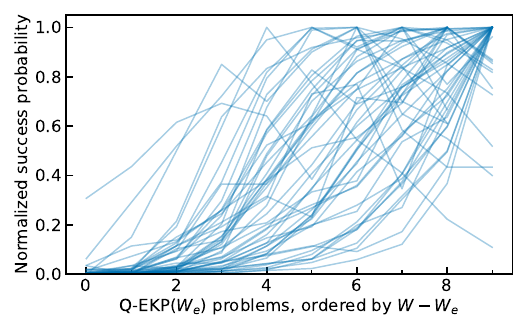}
    \caption{Same as \cref{fig:NormalizedSuccessProbabilitiesQAOA}, but for the D-Wave device.}
    \label{fig:NormalizedSuccessProbabilitiesDWave}
\end{figure}

In this appendix we consider the performance of the slackless method across the problems $\QEKP[W_e] \in \mathcal{Q}$ for several problem instances. In \cref{fig:NormalizedSuccessProbabilitiesQAOA}, we show the probability of sampling $\mathbf{x}^{*}$ across $\mathcal{Q}$ for all \KP problem instances with $n=16$, when using LR-QAOA with $p=10$. Since the absolute success probabilities vary across instances, we normalize the probabilities to lie in the same range. Note also that each problem instance may have different $w_{\max}$, leading to a different size for the set $\mathcal{Q}$ (see \cref{subsec:Implementation}). Consequently, the curves in \cref{fig:NormalizedSuccessProbabilitiesQAOA} terminate at different points along the sequence of \QEKP[W_e] problems.

For every problem instance in \cref{fig:NormalizedSuccessProbabilitiesQAOA} the same pattern is observed: the probability of sampling $\mathbf{x}^{*}$ increases as $W_e$ decreases. Thus, the success probability curve shown for an example instance in \cref{fig:SlacklessMethodsQAOA}(b) seems to be representative for other problem instances of the same size. The average success probability of the slackless method benefits from the higher success probabilities of the problems \QEKP[W_e] with smaller $W_e$. We note in passing that, at least for LR-QAOA with $p = 10$, this observation suggests a possible strategy for the slackless method: solving only the problems $\QEKP[W_e]$ with smaller $W_e$, as they are more likely to yield $\mathbf{x}^{*}$. This reduces the number of problems that need to be solved.

In \cref{fig:NormalizedSuccessProbabilitiesDWave}, the probability of sampling $\mathbf{x}^{*}$ across $\mathcal{Q}$ for all \KP problem instances with $n=16$ is shown for the D-Wave device. The success probabilities are again normalized. In contrast to the LR-QAOA results, the probability of sampling $\mathbf{x}^{*}$ does not consistently increase as $W_e$ decreases across all problem instances. While this pattern is still observed for some instances, no clear dependence on $W_e$ is seen for others. 

\bibliographystyle{apsrev4-2}
\bibliography{Knapsack_paper}

\end{document}